\PassOptionsToPackage{table,xcdraw}{xcolor}

\documentclass[sn-mathphys]{sn-jnl}

\usepackage{booktabs}
\usepackage{multirow}
\usepackage[normalem]{ulem}
\useunder{\uline}{\ul}{}

\usepackage[export]{adjustbox}

\graphicspath{{Figs_introduction/}, {Figs_th_framework/}, {Figs_percolation/}, {Figs_network_dismantling/}, {Figs_cascading_failures/}, {Figs_response/}, {Figs_conclusions/}}

\jyear{2024}%

\usepackage{subcaption}

\theoremstyle{thmstyleone}%

\usepackage{amssymb}
\begin{document}

\title[Robustness and resilience of complex networks]{Robustness and resilience of complex networks}

\author[1,2,3]{\fnm{Oriol} \sur{Artime}}\email{oartime@ub.edu}
\equalcont{These authors contributed equally to this work.}
\author[4]{\fnm{Marco} \sur{Grassia}}\email{marco.grassia@unict.it}
\equalcont{These authors contributed equally to this work.}

\author*[5,6,7]{\fnm{Manlio} \sur{De Domenico}}\email{manlio.dedomenico@unipd.it}

\author[8]{\fnm{James P.} \sur{Gleeson}}\email{james.gleeson@ul.ie}

\author[9]{\fnm{Hern\'an A.} \sur{Makse}}\email{hmakse@ccny.cuny.edu}

\author*[4]{\fnm{Giuseppe} \sur{Mangioni}}\email{giuseppe.mangioni@unict.it}

\author[10,11,12,13]{\fnm{Matja{\v z}} \sur{Perc}}\email{matjaz.perc@um.si}

\author[14]{\fnm{Filippo} \sur{Radicchi}}\email{filiradi@indiana.edu}

\affil[1]{\orgdiv{Departament de Física de la Matèria Condensada}, \orgname{University of Barcelona}, \orgaddress{\street{Martí Franquès, 1}, \city{Barcelona}, \postcode{08028}, \country{Spain}}}

\affil[2]{\orgdiv{Universitat de Barcelona Institute of Complex Systems (UBICS)}, \orgname{University of Barcelona}, \orgaddress{\city{Barcelona}, \country{Spain}}}

\affil[3]{\orgname{Universitat de les Illes Balears}, \orgaddress{\street{Campus UIB}, \city{Palma}, \postcode{07122}, \country{Spain}}}

\affil[4]{\orgdiv{Department of Electric, Electronic and Computer Engineering}, \orgname{University of Catania}, \orgaddress{\street{V.le A. Doria, 6}, \city{Catania}, \postcode{95125}, \country{Italy}}}

\affil[5]{\orgdiv{Department of Physics and Astronomy}, \orgname{University of Padua}, \orgaddress{\street{Via F. Marzolo, 8}, \city{Padova (PD)}, \postcode{35131}, \country{Italy}}}

\affil[6]{\orgdiv{Padua Center for Network Medicine}, \orgname{University of Padua}, \orgaddress{\street{Via F. Marzolo, 8}, \city{Padova (PD)}, \postcode{35131}, \country{Italy}}}

\affil[7]{\orgdiv{Istituto Nazionale di Fisica Nucleare}, \orgname{Sez. Padova}, \orgaddress{\street{Via F. Marzolo, 8}, \city{Padova (PD)}, \postcode{35131}, \country{Italy}}}

\affil[8]{\orgdiv{MACSI, Department Of Mathematics \& Statistics}, \orgname{University of Limerick}, \orgaddress{\city{Limerick}, \postcode{V94 T9PX}, \country{Ireland}}}

\affil[9]{\orgdiv{Levich Institute and Physics Dept.}, \orgname{City College of New York}, \orgaddress{\street{160 Convent Ave}, \city{New York}, \postcode{NY 10031}, \state{New York}, \country{USA}}}

\affil[10]{\orgdiv{Faculty of Natural Sciences and Mathematics}, \orgname{University of Maribor}, \orgaddress{\street{Koro{\v s}ka cesta 160}, \city{Maribor}, \postcode{2000}, \country{Slovenia}}}

\affil[11]{\orgdiv{Department of Physics}, \orgname{Kyung Hee University}, \orgaddress{\street{26 Kyungheedae-ro, Dongdaemun-gu}, \city{Seoul}, \postcode{02447}, \country{Republic of Korea}}}

\affil[12]{\orgdiv{Department of Medical Research, China Medical University Hospital}, \orgname{China Medical University}, \orgaddress{\street{No. 2, Yude Rd.}, \city{Taichung}, \postcode{404332}, \country{Taiwan}}}

\affil[13]{\orgname{Complexity Science Hub Vienna}, \orgaddress{\street{Josefst{\"a}dterstra{\ss}e 39}, \city{Vienna}, \postcode{1080}, \country{Austria}}}

\affil[14]{\orgdiv{Center for Complex Networks and Systems Research, Luddy School of Informatics, Computing, and Engineering}, \orgname{Indiana University}, \orgaddress{\street{Luddy Center for Artificial Intelligence, Room 2032, 1015 East 11th St.}, \city{Bloomington}, \postcode{IN 47408}, \state{Indiana}, \country{SA}}}


\abstract{
Complex networks are ubiquitous: a cell, the human brain, a group of
people and the Internet are all examples of interconnected many-body
systems characterized by macroscopic properties that cannot be
trivially deduced from those of their microscopic constituents. Such
systems are exposed to both internal, localized, failures and external
disturbances or perturbations. Owing to their interconnected
structure, complex systems might be severely degraded, to the
point of disintegration or systemic dysfunction. Examples include
cascading failures, triggered by an initially localized overload in power
systems, and the critical slowing downs of ecosystems which can be
driven towards extinction. In recent years, this general phenomenon
has been investigated by framing localized and systemic failures in
terms of perturbations that can alter the function of a system. We
capitalize on this mathematical framework to review theoretical and
computational approaches to characterize robustness and resilience
of complex networks. We discuss recent approaches to mitigate the
impact of perturbations in terms of designing robustness, identifying
early-warning signals and adapting responses. In terms of applications,
we compare the performance of the state-of-the-art dismantling
techniques, highlighting their optimal range of applicability for
practical problems, and provide a repository with ready-to-use scripts,
a much-needed tool set.
}


\maketitle

\section{Introduction}

A broad spectrum of biological, social, socio-ecological and engineering systems is characterized by units -- e.g., proteins or neurons, individuals, species -- that exchange information by means of complex network of interactions~\cite{albert2002statistical,newman2003structure,boccaletti2006complex}. The resulting connectivity patterns are usually heterogeneous and fat-tailed, with a few units acting as hubs and several other units with a few connections~\cite{barabasi1999emergence,broido2019scale,gerlach2019testing,voitalov2019scale,serafino2021true}, exhibiting a marked mesoscale organization into modules~\cite{guimera2005functional,newman2012communities,fortunato2016community,peixoto2017modelling,fortunato2022}, hierarchies~\cite{ravasz2003hierarchical,clauset2008hierarchical,peixoto2014hierarchical}, networks of networks~\cite{buldyrev2010catastrophic,mucha2010community,gao2012networks,de2013mathematical,artime2022multilayer}, high-order structures~\cite{lambiotte2019networks,battiston2021physics,bianconi2021higher} and latent geometry~\cite{de2017diffusion,garcia2018multiscale,boguna2021network}

Given the pervasive presence of complex networks in natural and artificial systems, it is of paramount importance to understand under which conditions they can be fully functional and to which extent they are fragile to external perturbations or internal failures~\cite{albert2000error}. In fact, the interconnected nature of complex networks can be either beneficial or detrimental to disturbances, exhibiting a rich behavior that ranges from absorbing -- without relevant large-scale effects -- specific types of perturbations to amplifying -- at system level -- microscopic disruptions. An emblematic example is the unfolding of a cascade of failures originated by perturbations localized on nodes which play a central role for information exchange -- e.g., signalling, electricity redistribution, etc -- between system's units~\cite{motter2002cascade,motter2004cascade}. Remarkably, the robustness and the resilience of complex networks~\cite{dorogovtsev2008critical, liu2022network} is characterized by different types of phase transitions, depending on the system's structural features -- e.g., spatially embedded or not, interdependent with other systems or not, multiplex, so forth and so on~\cite{bashan2013extreme,zhao2016spatio,radicchi2017redundant} -- and dynamics -- e.g., regimes where collective behavior emerges such as in synchronization, epidemic spreading or coupled dynamics~\cite{arenas2008synchronization,pastor2015epidemic,de2016physics,o2017oscillators}.

The study of system's response to perturbations in terms of structural and dynamical stability is crucial for applications, since it can be used to anticipate critical transitions~\cite{scheffer2012anticipating}. For instance, understanding cascading failure and robustness of a cell metabolic network~\cite{guimera2005functional,smart2008cascading} to activation or inhibition of specific enzymatic reactions can be used for developing specific therapies, drug repurposing and, more broadly, network and system's medicine~\cite{barabasi2011network}. Similarly, the robustness of protein-protein interaction networks to internal errors or external disruptions is strictly related to cell's function and its resilience across the tree of life~\cite{zitnik2019evolution}. In the human brain~\cite{bullmore2009complex}, localized disruptions can be related to strokes~\cite{siegel2016disruptions} and other pathological conditions. The stability~\cite{may1972will,holling1973resilience,pimm1984complexity} and restoration~\cite{pocock2012robustness} of ecological and socio-ecological~\cite{bascompte2009assembly,baggio2016multiplex} systems depend on their robustness to disturbances. In social systems, the containment of epidemic outbreaks is intimately related to the percolation features of the underlying social network~\cite{pastor2015epidemic}, while the propagation of systemic risks in financial and economic systems~\cite{gai2010contagion,cimini2015systemic,bardoscia2021physics,grassia2022insights} depends on their resilience to cascading failures, similarly to what happens in power grids which leads to generalized blackouts~\cite{carreras2002critical,yang2017small}. The efficiency~\cite{crucitti2003efficiency,bertagnolli2021quantifying} of traffic flows, from communication systems such as the Internet~\cite{doyle2005robust,de2017modeling} to transportation networks such as the air routes~\cite{scott2006network}, and their tolerance to errors and disruptions can be understood from percolation  analysis.

This is a non-exhaustive list of successful applications. Nevertheless, despite their ubiquity and the importance of understanding the conditions leading to systemic breakdown, a systematic overview of the literature is still missing. We fill this gap by reviewing existing protocols for network dismantling and classify empirical case studies into three main phases: i) designing of robustness; ii) early-warning signals; iii) adaptive responses. Specifically, in Sec.~\ref{sec:th_framework} we introduce the theoretical framework to operationally define robustness and resilience. In Sec.~\ref{sec:percolation} we review the connection with percolation theory, while in Sec.~\ref{sec:network_dismantling} we describe theoretical and computational techniques adopted for optimal percolation and network dismantling. Section~\ref{sec:cascading_failures} is devoted to describing cascading failures and the mechanisms leading to phase transitions that characterize the propagation of systemic risks. In Sec.~\ref{sec:response} we describe the methods to prevent and react to systemic collapse and, finally, we conclude by identifying potential research directions for the next future.

\section{Modeling robustness and resilience in interconnected systems}
\label{sec:th_framework}

Throughout this review, we will consider systems characterised by a network structure. A network is a collection of units (nodes) non-trivially connected with each other by means of edges or links. In the physics literature it is also common to describe them in terms of sites and bonds. A network can be mathematically represented by an adjacency matrix $A_{ij}$ where entries are positive if there is an interaction or a relationship between the pair of nodes $i$ and $j$, or are zero otherwise. At the microscopic level, a widely used measure of connectivity is given by node's degree $k_i$, quantifying the number of links involving node $i$, while more sophisticated descriptors can be used to capture a variety of features, from the tendency to cluster in triangles to the centrality in information exchange between units~\cite{boccaletti2006complex}.

Understanding how the structure and function of a network is affected by the failure of individual elements (i.e., nodes and/or edges) is a challenging task. In fact, microscopic failures do not sum up linearly and, while most failures may not heavily affect the functionality of the underlying system, the removal of specific elements may cause its collapse. The more abrupt the transition to a dysfunctional state, the more challenging to capture early-warning signals to prevent it or to devise effective responses to mitigate it.

There are a variety of scenarios that should be considered, each one with very different implications.
On the one hand, one can assume that internal failures are randomly distributed, caused by a node (e.g., a router) or an edge (e.g., a communication channel) breaking down. Yet, most real-world networks have shown high robustness to such random failures~\cite{albert2000error}, thanks to their highly heterogeneous connectivity~\cite{barabasi1999emergence}.
On the other hand, network structure can be exploited by an agent to intentionally break the system via targeted attacks driven by some protocol. This is the case, for instance, of immunisation policies~\cite{PhysRevLett.117.208301}, the crackdown of criminal, misinformation or malware networks~\cite{10.1093/comnet/cny002,Ren6554}, but also malicious attacks to power and natural gas distribution systems~\cite{SciGRIDv0.2,grassia2021machine}. Note that, however, each attack has an intrinsic cost that depends on the specific system, since removing a node or an edge translates into some kind of empirical action, such as vaccination, arrest, shutdown of a server, so forth and so on. An agent aims to minimising such a cost by removing the minimum number of nodes and edges needed to reach a given target. In mathematical terms, this procedure is translated into the design of a removal protocol and a cost function to be optimised accordingly.

Unfortunately, finding the optimal set of sites or bonds to target is a challenging task even on small networks.
An exhaustive search would need to explore the removal of all the possible combinations of nodes or edges, an operation that scales at least exponentially with system's size $N$, 
since the number of units to removal needed for disrupting the system is not known \emph{a priori}~\cite{Braunstein12368}.
For instance, finding the optimal set of nodes on a small network with $50$ nodes would require testing $2^{50} \approx 10^{15}$ combinations. 
The corresponding combinatorial optimisation problem, called Network Dismantling, is thus NP-hard and has usually been approached in rather different ways, including approximate theoretical models and computational heuristics. 

Furthermore, there is no general agreement in the literature on how to measure the health state of a system, or on what is the dismantling target, i.e., the goal of the attack.
That is, 1) how do we quantify the damage done so far to the system in order to stop the attack? And, 2) what function should we try to optimise during the dismantling process?
Regarding the first point, the most common approach~\cite{doi:10.1073/pnas.1009440108} involves percolation-related metrics, like monitoring the size of the Largest Connected Component (LCC) and stopping the attack once it has reached a given target size, although other network metrics such as efficiency~\cite{crucitti2003efficiency,kinney2005modeling} or nestedness~\cite{bascompte2009assembly,pocock2012robustness, alves2019nested} have been used. Yet, we should note that there is also no consensus on how small should the connected components be after the dismantling process, since it is application specific.
In other words, what size (either relative or absolute) can be associated with the failure of the underlying system strongly depends on the system itself and on the application.
For this purpose, some common values are $1\%$, $10\%$ and $18\%$ and $80\%$ of the original size~\cite{cohen2001breakdown,grassia2021machine,barabasi1999emergence,Braunstein12368,Ren6554}.
Regarding the second point, the optimization goal, while some works aim at finding the smallest set of nodes possible~\cite{albert2000error,Braunstein12368}, others aim at reducing the size of the Largest Connected Component as much as possible since the beginning of the attack~\cite{grassia2021machine}, whereas others to reduce the dismantling cost~\cite{Ren6554} as they assume that taking down some nodes is more expensive than the others.
Of course, it is easy to see that such goals are not necessarily compatible, and that the choice of the goal depends again on the specific application and on the available resources (e.g., time or money) that can be employed during the attack.

Another issue concerns the computational cost of the attacking algorithm itself, which can be measured in terms of run time and memory usage.
Computational (or time) complexity~\cite{cormen2022introduction} offers a way to measure the time spent by an algorithm to solve a problem, and thus can be used to compare different algorithms rigorously.
Specifically, the main assumption is that the run-time of the algorithm is proportional to the number of elementary operations it performs (expressed as a function of the input size), and that any kind of operation takes the same amount of time.
While this assumption is not exactly true, it is a good approximation that measures how the time spent scales with the input size, and that is not affected by the specific configuration of hardware, software, or programming language, since --- at least theoretically --- one could build an optimized configuration for each algorithm.
In particular, the most used approximation, commonly used for the comparisons, is the Big-O notation~\cite{cormen2022introduction}, representing the asymptotic number of operations performed as a function of the input size, and usually refers to the worst-case scenario, unless expressly specified otherwise.
In fact, the best case is not relevant, since it is usually associated with trivial instances, and the average case is generally hard to compute, since it depends on the specific input and path in the algorithm.
As an example, a time complexity of $O(N)$ indicates that, given the size of the input $N$, the algorithm will perform $N$ operations, while $O(N^2)$ indicates that it will perform $N^2$ operations.
Similar considerations can be made for the spatial complexity, which measures how the memory usage scales with the size of the input.
However, it is worth noting that the spatial complexity is usually not considered in the literature, since it is usually not a limiting factor for the algorithms.

In the next section, we will discuss a direction connection between the aforementioned theoretical problems and percolation theory, a framework widely used in statistical physics to study the behavior of a system and its critical response to perturbations.

\section{Connection with Percolation Theory}\label{sec:percolation}

\begin{figure}[h]
    \centering
    \includegraphics[width=0.95\linewidth]{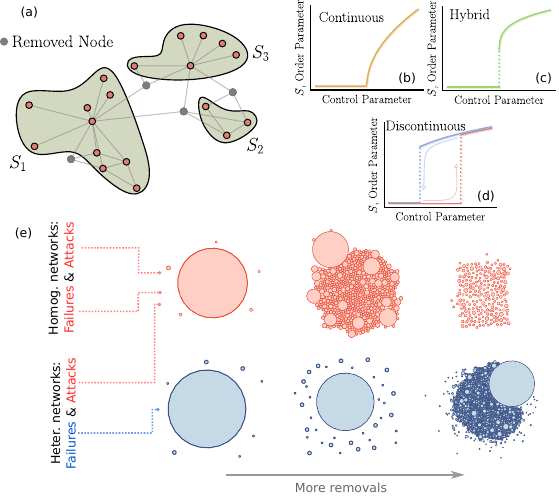}
    \caption{\textbf{Percolation as static approach to network robustness.} (a) Sketch of a network disintegration due to the removal of subsets of nodes, chosen according to a predefined protocol of removal $\phi$. Node removal predominates in this review, yet most of the concepts, metrics and techniques discussed throughout can be easily framed for link removal. The nature of $\phi$, related to the control parameter, heavily impacts on the response of the network, resulting in different sorts of phase transitions (b-d) for the size of the giant (largest) component, see the main text for details. In (e), we sketch how networks with homogeneous and heterogeneous connectivity patterns (roughly speaking, degree distributions with a second moment of similar order to the square of the mean or much larger than the mean, respectively) respond to failures and targeted attacks. Circles represent isolated components whose radius depend on their number of nodes. Each vertical column assumes the same number of nodes/links removed. Parts of the panel (e) are reproduced from Ref.~\cite{artime2020effectiveness}.}
    \label{fig:FigPercolation}
\end{figure}

Percolation is arguably the most direct and intuitive framework to approach network dismantling and, consequently, the most studied one. It is a famous model theoretically introduced in the study of gelation~\cite{flory1941molecular}, later developed under the umbrella of statistical physics~\cite{stauffer2018introduction, dorogovtsev2008critical} and probability theory~\cite{broadbent1957percolation}, and that has found many applications in different areas of science and engineering~\cite{isichenko1992percolation, sahini1994applications, araujo2014recent}. Percolation can be thought as an experiment consisting in removing nodes and/or links according to some predefined rules (also named attacking protocols), to compute different statistical and geometrical properties of the residual network, as schematically shown in Fig.~\ref{fig:FigPercolation}(a). This approach allows one to track the structural response of a system while its components are, for some reason (failures, maintenance, attacks, etc.), ruled out. The state of these components is considered binary: present (functional) or removed (failed), and one usually computes metrics related to the sizes of the remaining connected components formed by the non-failed nodes. The fruitful connection between percolation theory and dismantling originates from the assumption that a bare-bones requirement for a system to function properly, whatever its function, is to be globally connected. Hence, the loss of functionality due to the network degradation maps to the phase transition of the percolation model. Operationally, this is conveniently characterised by quantities such as the critical point and the size of the giant component, offering quantitative and qualitative insights of the dismantling process, see Figs.~\ref{fig:FigPercolation}(b-d).

In complex networks, nodes are not structurally or functionally equivalent as it happens in regular lattices. There exist a plethora of methods and metrics to rank them according to different criteria, which are usually application driven. This intrinsic heterogeneity offers a wide spectrum of attacking protocols, allowing one to assess network robustness under many physically relevant scenarios. For instance, a random selection of nodes is used to model failures and errors occurring in a system, while targeted attacks -- driven by a more sophisticated protocol -- can be implemented as informed interventions. The latter includes making use of topological information, such as removing the nodes with the largest number of connections or with largest centrality~\cite{rodrigues2019network, holme2002attack}, or non-topological information, such as node or edge metadata~\cite{artime2021percolation}. 

There is a non-trivial interplay between the dismantling protocol and network topology. Pioneering studies~\cite{albert2000error} numerically showed that, in both synthetic and empirical networks, homogeneous or heterogeneous connectivity patterns can induce radically different responses (Fig.~\ref{fig:FigPercolation}(e)). The former reacts to failures and attacks in a similar way since the largest possible degree is not far from the mean value, yielding always a finite dismantling point (upper row in Fig.~\ref{fig:FigPercolation}(e)). This is not true, though, for the latter, such as scale-free networks, because of the role played by the hubs: they are extremely good at holding together the network when malfunctions are placed uniformly at random, but if they are directly targeted then the network breaks apart very quickly -- the so-called ``robust-yet-fragile'' property of heterogeneous networks, see the rightmost column of Fig.~\ref{fig:FigPercolation}(e). This agrees with the prediction of dismantling points in uncorrelated networks~\cite{molloy1995critical, cohen2002percolation}, and it was further formalised and quantified theoretically by using an approach based on generating functions~\cite{gordon1962good, newman2001random, callaway2000network, gross2022network}. The advantage of the methods that rely on generating functions is their mathematical flexibility to encompass generalisations, while they offer some predictive power, e.g., about critical exponents~\cite{cohen2002percolation, moore2000exact, goltsev2008percolation} or closed-form equations for the critical point and the size of the giant component~\cite{newman2018networks}. Such methods assume that the underlying network is an annealed version of the configurational model, hence they tackle robustness at an ensemble level and become useful to unravel the role played by the functional form of the degree distribution $p_k$ or its parameters. For instance, given a degree-dependent protocol of node removal $\phi_k$, the size of the giant component $S^{\text{GF}}$ is given by the solutions of
\begin{align}
    & S^{\text{GF}} = G(1) - G(u) \\
    & u = 1 - H(1) + H(u),
\end{align}
where $G(z) \equiv \sum_k p_k \phi_k z^k $ and $H(z) \equiv \frac{1}{\langle k \rangle} \sum_k (k+1) p_{k+1} \phi_k z^k $ are the generating function of the degree distribution and the excess degree distribution, respectively. $\langle k \rangle$ denotes the mean degree. The dismantling point is given by $H'(1) = 1$. 

Similar equations can be written for the case of random link failures~\cite{li2021percolation}. It immediately follows that the location of the dismantling point coincides in site and bond percolation. Remarkably, the critical exponents characterising the phase transition also coincide, and they are network-independent and equal to the mean-field predictions~\cite{stauffer2018introduction}, as long as the connectivity patterns in the network are not too heterogeneous. For instance, in scale-free networks there has been found a violation of this universality equivalence~\cite{radicchi2015breaking}, and the critical indices may depend on the exponent of the degree distribution~\cite{cohen2002percolation}. The case of targeted attacks on links has been less explored in the literature. To encode the link information in the dismantling process, one would need to consider simultaneously the topological properties of the two nodes at the end of the link, requiring a more involved mathematical treatment than the one presented above~\cite{shiraki2010cavity}. Despite of this fact, many empirical networks are weighted~\cite{barrat2004architecture} or have features/metadata defined on the edges, hence formalising the dismantling problem on these cases will surely be a fruitful future application of percolation theory.

The above methods might be not always practical to accurately predict the robustness of empirical networks, because one aims to know how a specific network topology, and not the underlying ensemble, reacts to malfunctions. In this case, message-passing methods, which take as input the actual connectivity of the networks instead of their degree distributions, might be more convenient to compute percolation quantities~\cite{hamilton2014tight, karrer2014percolation, radicchi2015percolation, newman2022message}. This leads to a more complicated mathematical and numerical treatment, but it often offers better estimates for the giant component and the critical point than the method based on generating functions~\cite{radicchi2015predicting, radicchi2016beyond}. As an example, if we denote by $\phi_i$ the probability that node $i$ has not been removed, the size of the giant component $S^{\text{MP}}$ of a network is given by
\begin{align}
    & S^{\text{MP}} = \frac{1}{N} \sum_{i=1}^N s_i \\
    & s_i = \phi_i \left[ 1 - \prod\limits_{j \in \mathcal{N}_i} \left( 1 - t_{i \to j} \right) \right] \label{eq:si} \\
    & t_{i \to j} = \phi_j \left[ 1 - \prod\limits_{k \in \mathcal{Q}_{i \to j}} \left( 1 - t_{j \to k}\right) \right]. \label{eq:tij}
\end{align}
Here, $s_i$ represents the probability of node $i$ to belong to the giant component, $\mathcal{N}_i$ denotes the neighborhood of node $i$ and $\mathcal{Q}_{i \to j}$ stands for the subset of nodes to whom $j$ \textit{passes} a (directed) \textit{message} regarding the probability of belonging to the giant component through the edge that joins them. Typically, $\mathcal{Q}_{i \to j} = \mathcal{N}_j \backslash \lbrace i \rbrace$ to avoid backtracking messages, although other choices that take short-range loops into account are possible~\cite{radicchi2016beyond}. Close to the dismantling point, all the $t_{i \to j}$ tend to $0$, so Eq.~\eqref{eq:tij} can be expanded to obtain $\vec{t} = \vec{\phi} \circ (\hat{G} \vec{t})$, where $ \circ $ is the Hadamard product and $\vec{\phi} \equiv \left( \phi_1, \ldots \phi_1, \ldots, \phi_N \ldots \phi_N \right)$, with each $\phi_i$ appearing $k_i$ times. $\hat{G}$ is a logical matrix of dimension $2 \lvert E \rvert \times 2 \lvert E \rvert$, where $\lvert E \rvert$ is the number of edges, and it depends on $\mathcal{Q}_{i \to j}$. The dismantling condition is given by the emergence of a non-trivial solution in the eigenvalue problem $\hat{G} \vec{t} = \vec{t} \oslash \vec{\phi} \equiv \lambda \vec{t}$, where $ \oslash $ is the Hadamard division and $\lambda$ is the eigenvalue. By virtue of the Perron-Frobenius theorem, we know the non-trivial solution is related to the largest eigenvalue of the operator $\hat{G}$. For a constant occupation probability $\phi_i = \phi$, one obtains $\phi_c = 1/\lambda_{\text{max}}$, where $\lambda_{\text{max}}$ is the largest eigenvalue of $\hat{G}$.

Message-passing methods can be developed for bond percolation too. If a fraction $1 - \phi$ of links chosen uniformly at random is removed from the network, then the size of the giant component is given by $S^{\text{MP}} / \phi$. Thus, message passing predicts the same percolation threshold for site and bond percolation, in agreement with the predictions based on generating functions~\cite{radicchi2015breaking}. Network dismantling based on link attacks can be easily implemented in this framework, since the removal probability of each individual link can be encoded in a new vector of occupation probabilities $\vec{\phi}'$ to be plugged in the message-passing equations for bond percolation.

Percolation theory also helps to make predictions on the robustness of networks that display topological correlations, a feature that characterises most empirical interconnected systems. One of these correlations is the so-called assortative mixing~\cite{newman2002assortative}, i.e., the tendency of nodes to be connected to peers of similar kind. It has been identified that networks that are assortative in their degree tend to be more robust than those with disassortative patterns because hubs create a redundant core that hold the network together when faced to both random and hub attacks~\cite{newman2003mixing}. The presence of local clustering, i.e., overabundance of closed triangles with respect to a randomised network, has been also considered in different ways. Depending on the way clustering is defined and on whether its value is high or low, the network might be considered more or less robust than its unclustered counterpart, see, e.g.,~\cite{serrano2006percolation, serrano2006clustering, berchenko2009emergence, newman2009random}. Going beyond local topological correlations, the robustness of networks displaying mesoscale nontrivial structures lends itself to be also addressed under the percolation framework. One is core-periphery network organisation~\cite{rombach2014core}. When these structures are attacked randomly they can display two dismantling points, one due to the nodes in the core and one for those in the periphery~\cite{colomer2014double, allard2017asymmetric, hebert2019smeared}. Another is $k$-cliques, whose percolation-based analysis has been useful to identify overlapping communities~\cite{derenyi2005clique}.

From everyday experience, we know dismantling can occur in an abrupt manner that is very difficult to anticipate due to the absence of apparent early-warning signals. Identifying those scenarios in which a sudden transitions might occur is extremely valuable since they can produce a huge socioeconomic impact. Percolation theory offers a diversity of mechanisms that can induce such abrupt topological fragmentations. One is the so-called $k$-core percolation, where all the nodes with degree less than $k$ are removed, iteratively~\cite{dorogovtsev2006k}. For $k = 2$ the transition is continuous, but for $k > 2$ it becomes hybrid. Bootstrap percolation, which allows removed nodes to recover if they have a number of neighbours above a certain threshold, can present also discontinuous and hybrid transitions~\cite{baxter2010bootstrap}. Additionally, there is a family of models based on selection rules that might considerably accelerate or delay the critical point, at the expenses, though, of drawing upon mesoscopic information~\cite{bohman2001avoiding, spencer2007birth, beveridge2007product, krivelevich2010hamiltonicity}. Among these, the most famous is the product rule that leads to the so-called explosive percolation~\cite{achlioptas2009explosive}. This product rule consists in picking uniformly at random two links and removing the one that maximises the product of sizes of the components it joins: in this way, the giant component collapses abruptly. Furthermore, the dismantling point occurs before than the percolation case based on random link picking. Despite of this abruptness, it was shown that the explosive transition of the product rule is actually continuous, but with a peculiar critical behaviour~\cite{riordan2011explosive, da2010explosive, grassberger2011explosive}. Further information about explosive percolation can be found in the recent review~\cite{d2019explosive}. Finally, a last mechanism that might yield to abrupt dismantling is associated to interdependencies when different networks are coupled. Interdependencies play a central role to the study of cascading failures, to be discussed in a next section, but they can be modelled with a static percolation framework and shed light on the conditions under which discontinuous transitions appear~\cite{son2012percolation}. 

All in all, percolation is a central model in network robustness thanks to its flexibility, its low computational cost and its analytical power~\cite{li2021percolation}. It proposes protocols to dismantle interconnected systems according to some metrics, thus unveiling the importance of these precise indicators on the system's vulnerability. However, in other cases, the intervening protocols are metric-agnostic and goal oriented. Following these lines, in the next section we focus on algorithms that look for strategies to dismantle networks as efficiently as possible.

\section{Optimal Percolation and Network Dismantling}\label{sec:network_dismantling}

Searching for the optimal strategy for the fastest network disintegration is the optimal percolation problem~\cite{morone2015influence}, also known as network dismantling~\cite{Braunstein12368}. This is a combinatorial optimisation problem that aims at finding the set of nodes (or edges) that fastest disrupts a network when removed, breaking the largest connected component into isolated sub-components and thus efficiently degrading the system's functionality.
Such an optimisation problem is computationally challenging even on small networks (i.e., it is NP-hard), and in the literature there is no agreement on how small the sub-components should be after the dismantling process, or even during the dismantling target.

\begin{figure}[htbp!]
	\centering
	\begin{subfigure}{1.\textwidth}
		\centering
		\includegraphics[width=0.9\textwidth,keepaspectratio,valign=c]{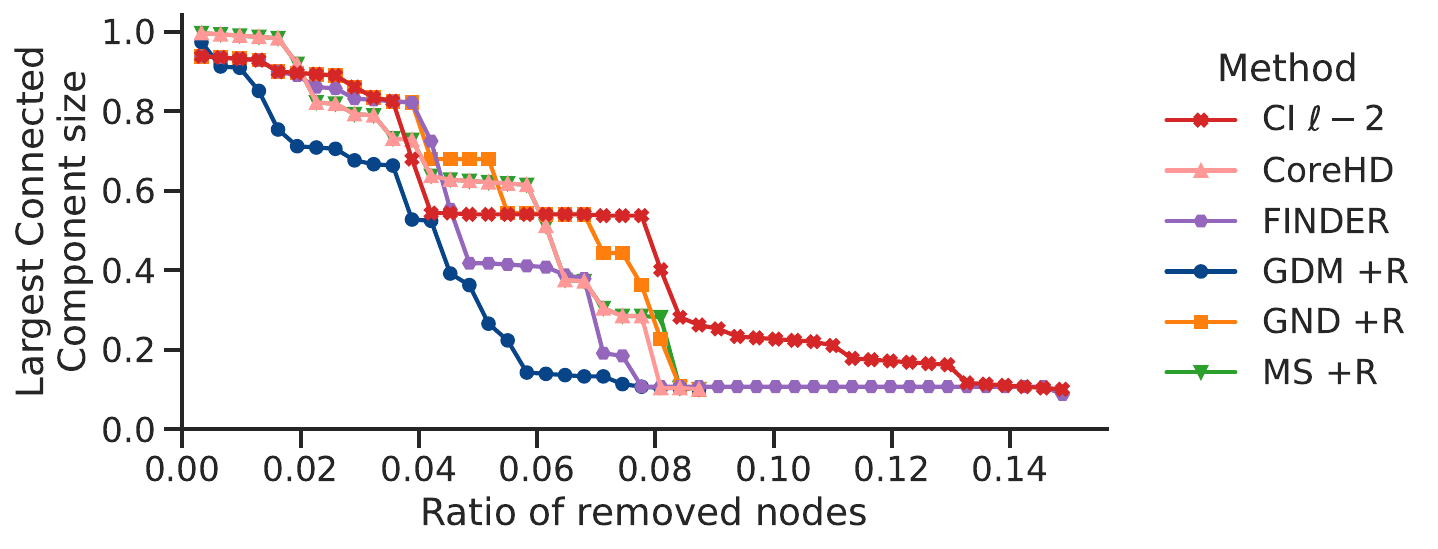}
		\caption{Dismantling process.}
		\label{f:corruption_dismantling}
	\end{subfigure}%
	\hfill
	\begin{subfigure}{0.33\textwidth}
		\centering
		\includegraphics[width=0.99\textwidth]{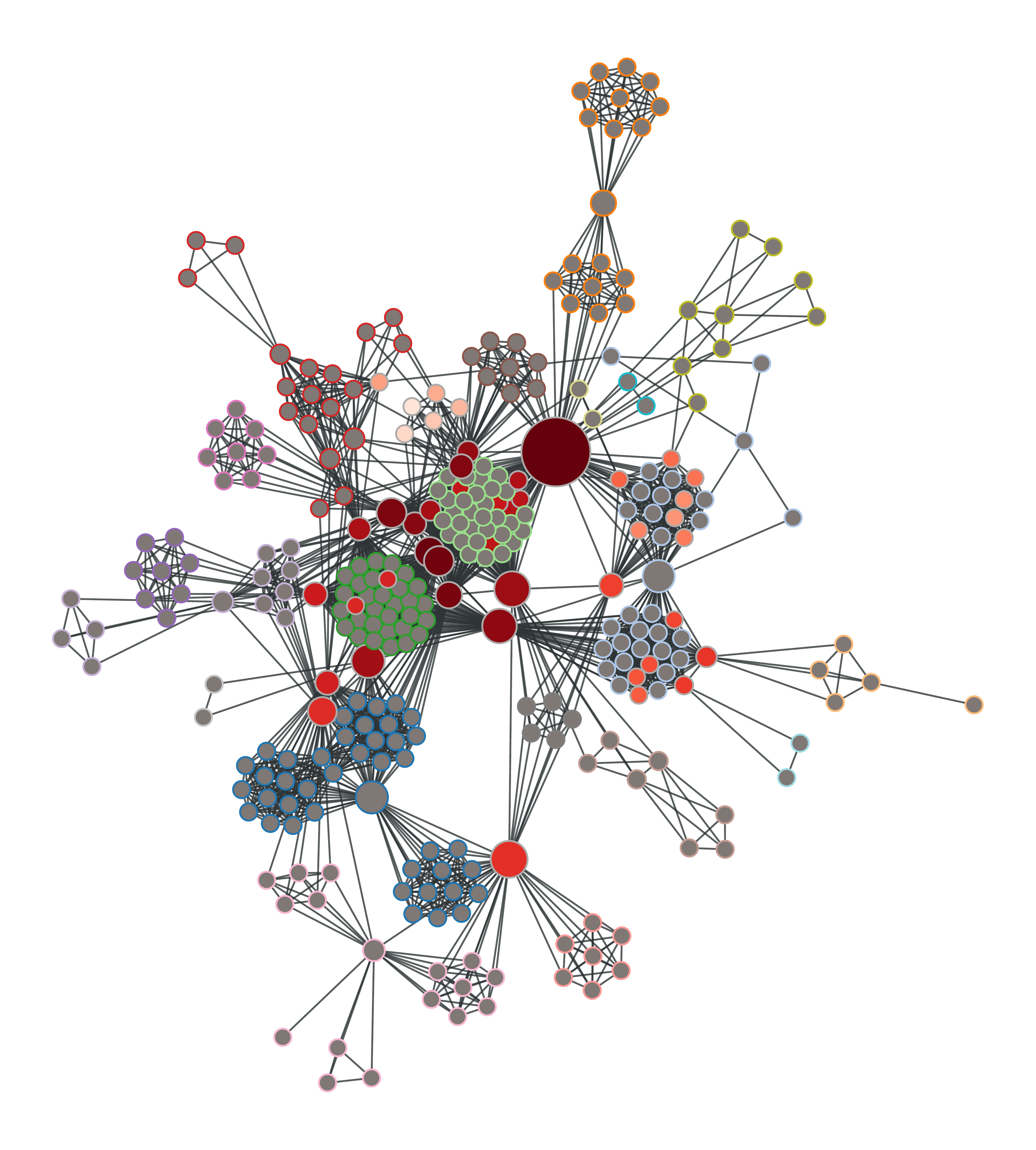}
		\caption{CI $\ell-2$}
		\label{f:attack_corruption_cil2}
	\end{subfigure}%
	\begin{subfigure}{0.33\textwidth}
		\centering
		\includegraphics[width=0.99\textwidth]{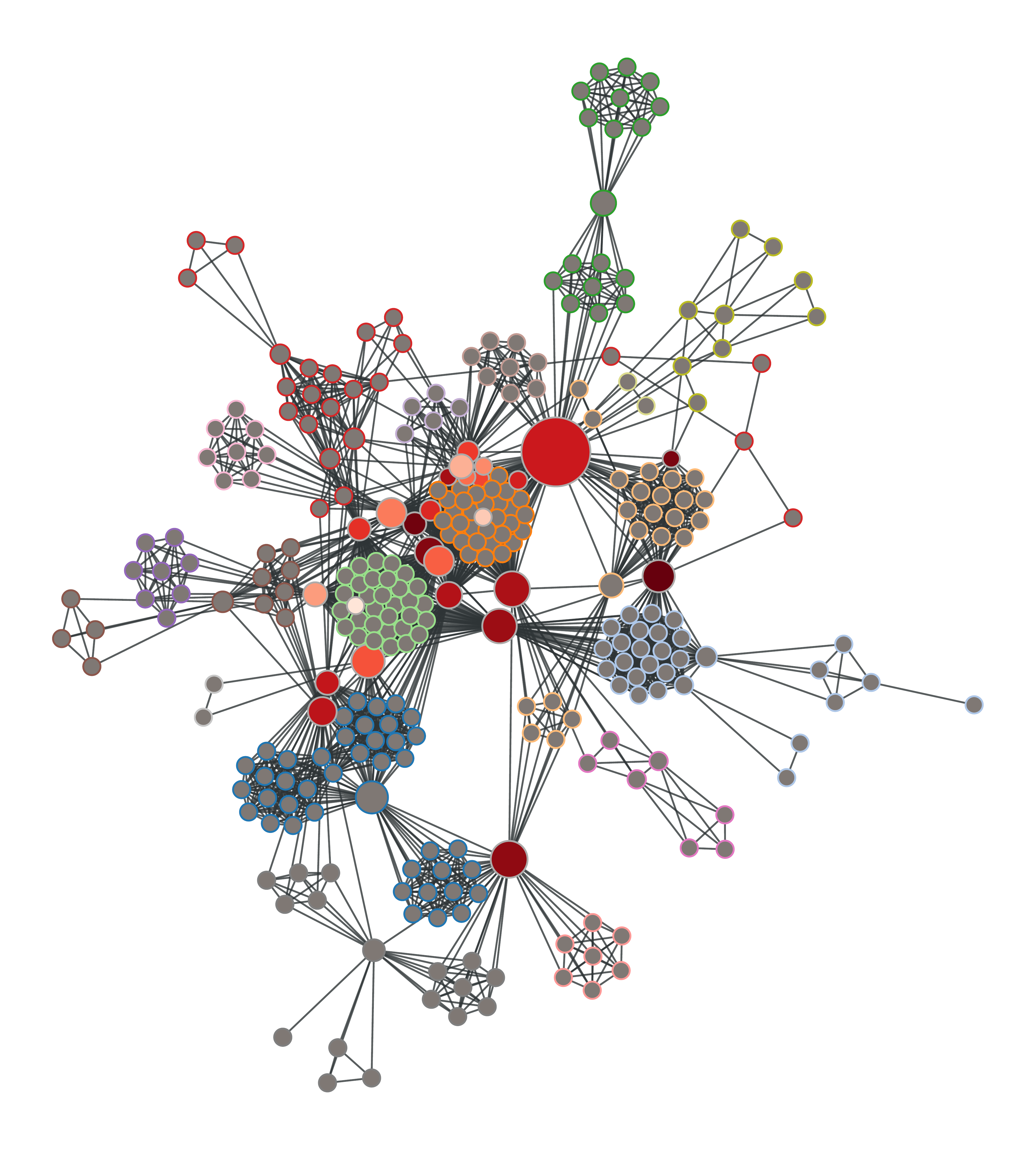}
		\caption{CoreHD}
		\label{f:attack_corruption_corehd}
	\end{subfigure}%
	\begin{subfigure}{0.33\textwidth}
		\centering
		\includegraphics[width=0.99\textwidth]{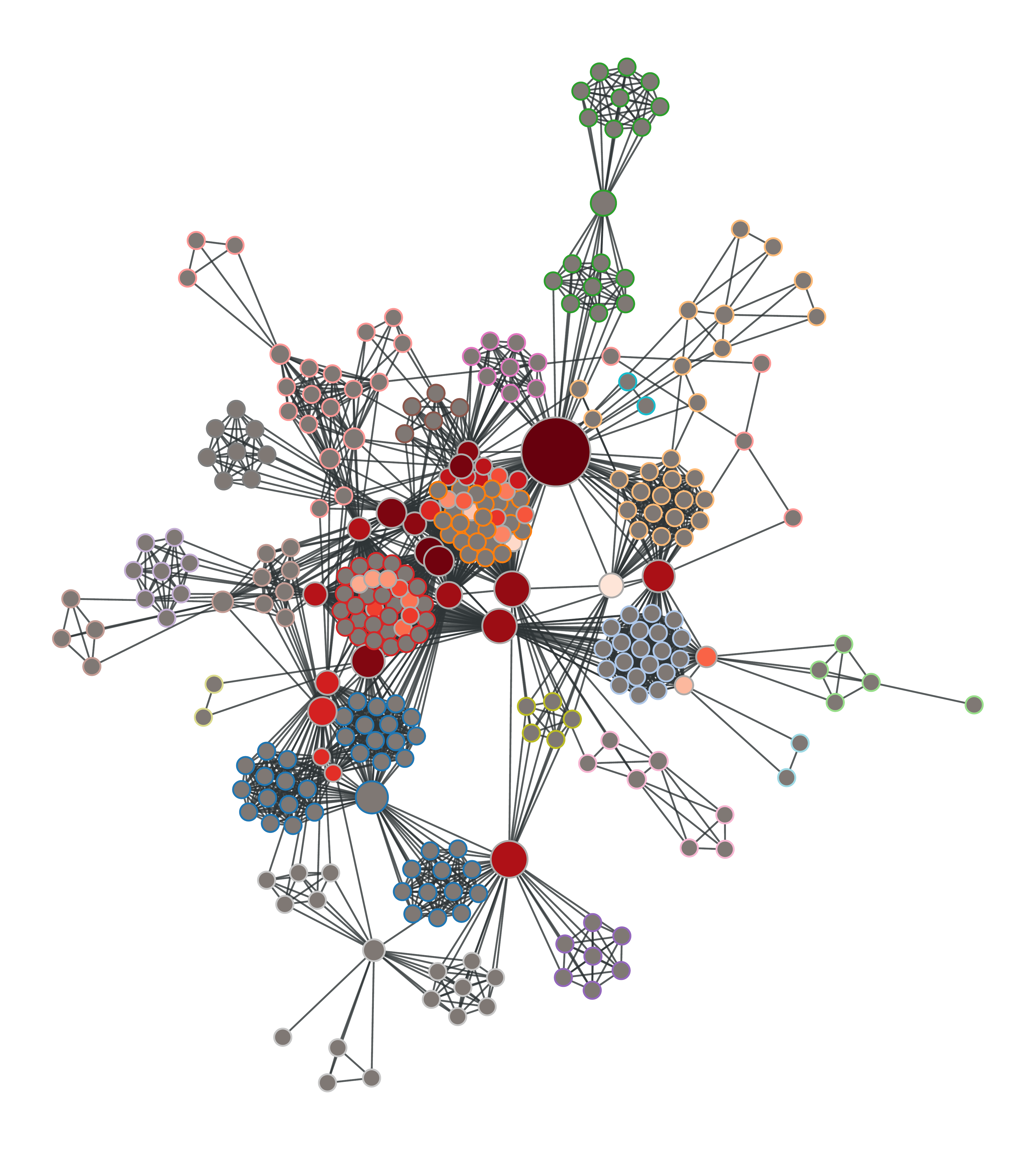}
		\caption{FINDER}
		\label{f:attack_corruption_finder}
	\end{subfigure}%
	\hfill
	\begin{subfigure}{0.33\textwidth}
		\centering
		\includegraphics[width=0.99\textwidth]{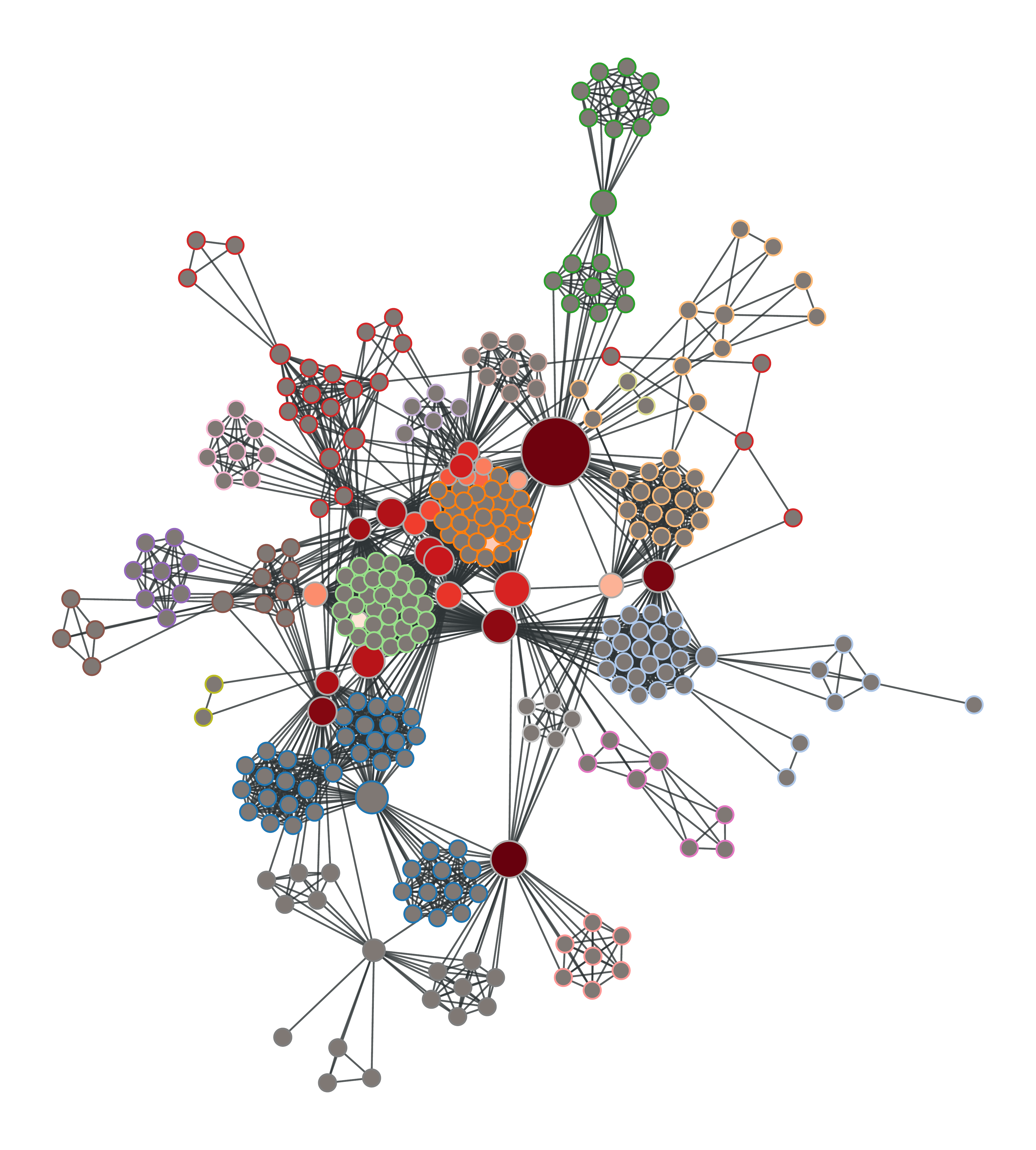}
		\caption{GDM+R}
		\label{f:attack_corruption_gdmr}
	\end{subfigure}%
	\begin{subfigure}{0.33\textwidth}
		\centering
		\includegraphics[width=0.99\textwidth]{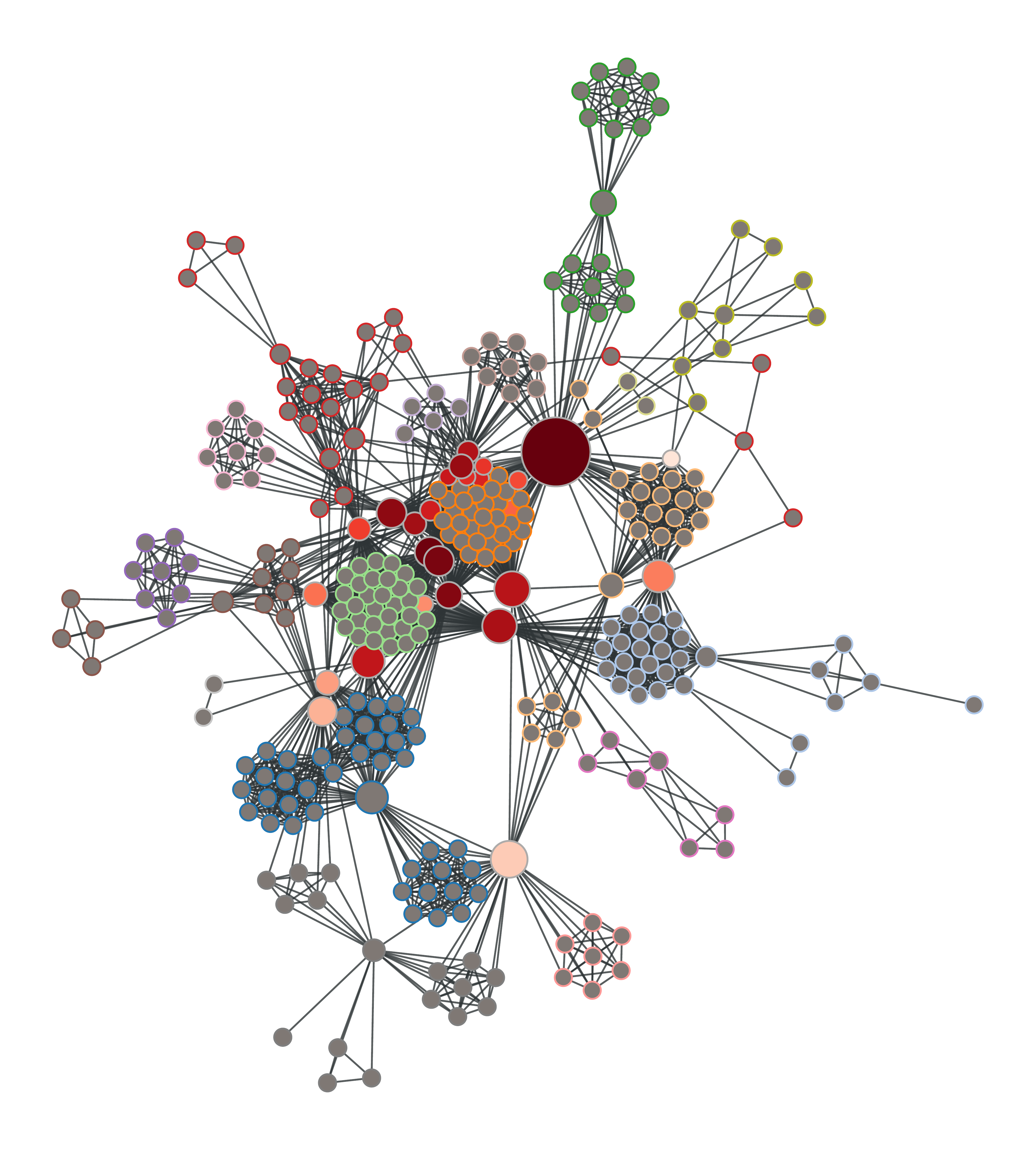}
		\caption{GND+R}
		\label{f:attack_corruption_gndr}
	\end{subfigure}%
	\begin{subfigure}{0.33\textwidth}
		\centering
		\includegraphics[width=0.99\textwidth]{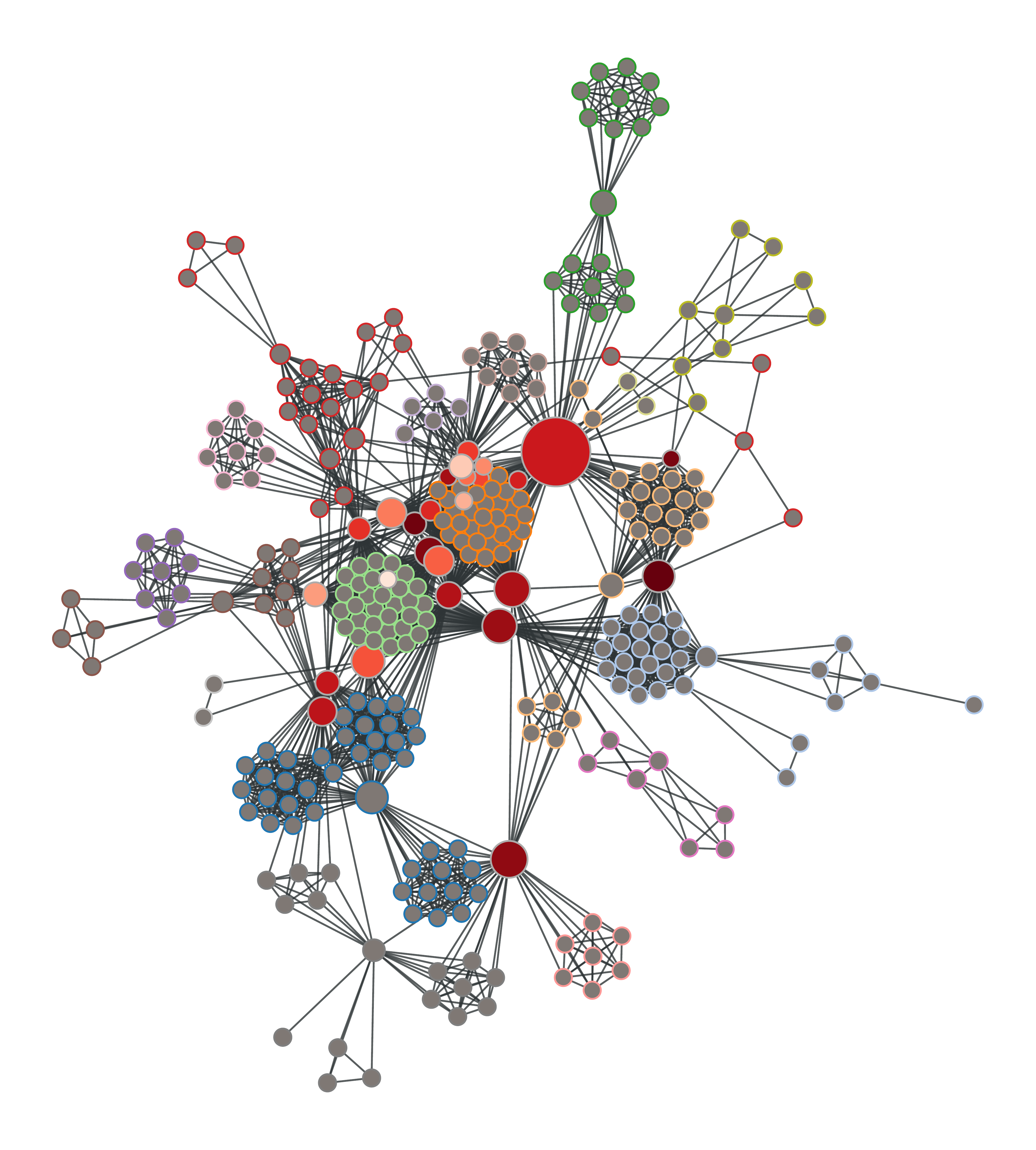}
		\caption{MS+R}
		\label{f:attack_corruption_msr}
	\end{subfigure}%
 \caption{\textbf{Comparison of state-of-the-art dismantling methods.} (a) The cutting-edge algorithms, described in Section~\ref{sec:network_dismantling} are compared in terms of their ability to drive the system -- a Brazilian corruption network~\cite{10.1093/comnet/cny002} with 309 nodes -- towards disintegration, as measured by the relative size of the largest connected component (LCC). In panels (b)--(g), we display the disintegrating paths of the different curves shown in (a). The color of the nodes represents (from dark red to white) the attack order (i.e., nodes in gray are not removed), while their size represents their betweenness value. The contour color of remaining nodes represents the cluster they belong to after the attack.}
 \label{f:corruption_dismantling_graph}
\end{figure}

Earlier work aimed at finding the smallest set of nodes to break the
network by characterising their importance by a series of topological
centralities. A simple strategy to break the network consists of using
a centrality score to rank nodes to ``attack'' the network. The most
basic is the degree of the node privileging the hubs, following the
intuition that the more connections the better
\cite{albert2000error,cohen2001breakdown}.
These strategies can also be divided in static and dynamic ones, the difference being that the former computes the removal order of nodes (or edges) at the beginning of the dismantling process, while the latter updates the order during it, thus accounting for the changing state of the network.

However, there are plausible circumstances where the crucial nodes do
not correspond to the most connected ones, but to those nodes that
have few connections, yet they are strategically located within the
core of the network. These topological bridges represent a fundamental notion in sociology termed by Granovetter as {\it ``The Strength of Weak Ties''}~\cite{granovetter1973}.  Many heuristic centralities have
been proposed to find these crucial nodes. They can be classified into
broad classes as: (i) degree-based (e.g., hubs and $k$-cores), (ii) shortest paths-based (e.g., closeness and betweenness centralities),
(iii) walks-based (e.g., eigenvector, eigenvalue and Katz
centralities), (iv) random walk-based (e.g., Pagerank), (v)
non-backtracking walk-based, and (vi) machine learning-based
approach. At a higher level of classification, we might distinguish between genuinely structural methods, based on topological descriptors, and genuinely dynamical methods, based on some flow dynamics between nodes.

Optimal percolation then systematises the search for the optimal set
by attempting to find a configuration of nodes $\mathbf{n}^*$
corresponding to the minimal fraction of removals $q_c^{\rm opt}$
such that the giant connected component of the network $G_{\infty}$ is
disintegrated optimally~\cite{morone2015influence}:
\begin{equation}
  q_c^{\rm opt}\,=\,\min \{ \, q \in[0,1] \,\, \vert \,\, G_{\infty}(q)=0\,\}\, .
\end{equation}

Optimal percolation is an NP-hard combinatorial problem~\cite{kempe2003maximizing}; intractable since an explicit functional
form of $G_{\infty}(\mathbf{n})$ is not feasible.  However, an
approximated solution for sparse networks was found in~\cite{morone2015influence} leading to the collective influence (CI)
algorithm~\cite{morone_2016}.  An approximate optimal set is obtained
as an (infinite) sequence of optimized attacks expressed as successive
approximations to the minimization of the largest eigenvalue of the
non-backtracking matrix. It leads to the CI index of node $i$ with
degree $k_i$:
\begin{equation}
\text{CI}_\ell(i) \ = \ (k_i-1)\sum_{j\in\partial\text{Ball}(i,\ell)}(k_j-1) ,
\label{eq:cia}
\end{equation}
where $\partial\mathrm{Ball}(i,\ell)$ is the surface of a ball of radius
$\ell$ -- estimated from shortest-path distances -- around node $i$. This algorithm gives a good
approximation to the optimal percolation set.  Later, better
algorithms based on message passing~\cite{1altarelli2014containing}
and belief propagation~\cite{1altarelli2013optimizing} have been
proposed, including decycling and dismantling~\cite{Braunstein12368,1mugisha2016identifying} and explosive
percolation~\cite{PhysRevLett.117.208301}.  They have enriched our
understanding by tackling the optimal percolation problem with
rigorous theories, and further generated a number of sophisticated and
efficient algorithms that are applicable to large-scale complex
systems.

Ref.~\cite{1mugisha2016identifying} showed that the optimal
disintegration of the giant connected component is achieved by
the disintegration of cycles found by the minimum feedback vertex set.
The rationale behind the connection between optimal percolation and
optimal decycling is that, for sparse random networks, short loops
rarely exist in small connected components. If the long loops in the
giant component are cut, the network will break into small tree
fragments. As indicated by Braunstein {\it et al.}
\cite{Braunstein12368}, the optimal decycling threshold $q_c^{\rm
  dec}$ acts as an upper bound of the optimal percolation threshold
$q_c^{\rm opt}$.  Optimal decycling is NP-hard and can be
approximately solved via belief propagation algorithms.  Two
approaches were developed, the Belief Propagation-guided Decimation (BPD)~\cite{1mugisha2016identifying} and the Min-Sum~\cite{Braunstein12368}
algorithms, that find smaller influencer sets than the Collective Influence algorithm,
although at the expense of increasing complexity.
In particular, both algorithms first decycle the network via message-passing algorithms, then break the resulting forest using an ad-hoc algorithm called tree-breaker.
While the computational complexity of each message-passing iteration is linear with the number of edges, the greedy tree-breaker algorithm proposed in~\cite{Braunstein12368} scales as $O(N (\log(N + T ))$, where $T$ is the maximal diameter of the trees inside the forest and $N$ is the number of nodes.
Furthermore, the Min-Sum algorithm also includes a reinsertion phase at the end of the attack, consisting in a greedy process that reintroduces the nodes that are not actually needed to reach the desired target.
The reinsertion phase is meant to reduce the number of nodes that are actually removed from the network --- thus decreasing the cost of the attack --- and should be used in heavy-tailed networks, where the decycling and the dismantling problems are not equivalent.
The Authors report a sub-linear time complexity for this phase, but the exact complexity is not known.

Inspired by these decycling-based algorithms, a simple and faster heuristic algorithm with complexity $O(N)$ (for sparse networks), CoreHD, was developed in~\cite{Zdeborov__2016}, the idea being to decycle the network by iteratively removing the nodes with the highest degree from the 2-core, then employ the same tree-breaking and reinsertion algorithms used by Min-Sum.
Explosive Immunization was used in~\cite{PhysRevLett.117.208301} based on explosive percolation transition featuring a discontinuous emergence of the giant component at the critical point.
Generalized Network Dismantling (GND) was proposed in~\cite{Ren6554,10.1007/978-3-030-36687-2_65} to address the non-unit cost of removals, and is based on iteratively removing those nodes that maximize an approximated spectral partitioning.

Other studies have also proposed Machine Learning-based attack strategies, like Graph Dismantling with Machine learning (GDM)~\cite{grassia2021machine}, and FINDER~\cite{fan2020finding}.
Both employ Geometric Deep Learning, specifically Graph Neural Networks (GNNs), to learn an attack strategy on small synthetic networks.
However, while GDM learns in a supervised manner on networks dismantled optimally via brute-force, FINDER learns in a reinforcement fashion.
Furthermore, CoreGDM, which is inspired by CoreHD as it attacks the 2-core of the networks using GDM models, was proposed in~\cite{10.1007/978-3-031-28276-8_8}.
Also, in~\cite{PhysRevResearch.5.013076} Authors investigate the performance of network embeddings in Euclidean and hyperbolic spaces.

Optimal percolation was also studied on multiplex networks~\cite{osat2017optimal} and game theory~\cite{szolnoki2016collective}.

In terms of dynamical information flow in the network, optimal
percolation aims to find the minimal set of superblockers that could
stop the spreading~\cite{chen2022influence}. A dual problem is to find the superspreaders of
information who maximize the spreading if selected as seeds. In general, these two problems are not necessarily equivalent as shown in~\cite{radicchi2017fundamental}. However, for a particular form of the
Linear Threshold Model of spreading with thresholds given by $k_i-1$,
it was shown~\cite{morone2015influence} that optimal percolation is
equivalent to the Influence Maximization Problem (IMP) introduced by
Kempe, Kleinberg and Tardos~\cite{kempe2003maximizing}. IMP consists
to find the optimal set of $k$-superspreaders (influencers) that would
initiate the largest scale propagation of information process in the
network. For a given number of initial spreaders $k=\lvert S\rvert$,
one wishes to find a $k$-node set $S$ of maximum influence $\sigma(S)$:
\begin{equation} \max_{\lvert S \rvert \le k} \sigma(S) \, .
\label{eq:imp}
\end{equation} 

This opens the applicability of optimal percolation to a wide variety
of societal problems and beyond~\cite{makse2023}. Theories of
influencers encounter applications for identifying influential
spreaders in social media, superspreaders of disease in pandemics,
essential gene mutations in genetic networks that could lead to
disease, essential areas in the brain for integration, keystone
species whose extinction could bring an ecosystem to the tipping point
of collapse, and financial institutions that are ‘too-big-to-fail’.

To this aim, it has been recently proposed an approach based on network states encoded into suitably defined density matrices~\cite{de2016spectral}, where robustness is analysed at multiple scales corresponding to the time scales required for information to diffuse through the network~\cite{ghavasieh2021unraveling}. Applications to empirical social, biological and transportation systems, have shown that nodes crucial for information dynamics are also responsible for keeping the network structurally integrated, while the opposite is not necessarily true and functional fragmentation happens before full structural disintegration~\cite{ghavasieh2023dismantling}.

A summary of the algorithms described above, along with their main features, can be found in Table~\ref{t:algorithms}. When performing robustness assessments one needs to evaluate which is the most physically meaningful and computationally efficient approach for each particular network to be dismantled. For instance, in Fig.~\ref{f:corruption_dismantling_graph} we show a quantitative and qualitative comparison between algorithms when applied to a corruption network. In this case, the GDM algorithm with reinsertion is the one that fastest dismantles it, in approximate 20 removals out of the 309 nodes. Yet, there is no one-size-fits-all algorithm, as we report in Table~\ref{t:real_world_results}. In there, we present a performance comparison on a large set of real-world networks from different application domains (biological, information, social and technological), finding that the best algorithm varies within and across domains. See Table~S1 and Table~S2 in SI for further details on the chosen empirical networks, and Tables~S3 and~S4 of the Supplementary Information for a comparison on synthetic networks and on the Lancichinetti-Fortunato-Radicchi (LFR)~\cite{PhysRevE.78.046110} model. 

\begin{table}[!ht]

  \centering
  \scriptsize
  \rowcolors{2}{gray!25}{white}
  \adjustbox{max width=\textwidth}{
    \begin{tabular}{@{}llllll@{}}
      \toprule
      Algorithm                   & Type                   & Static & Reinsertion & Computational Complexity & Ref. \\ \midrule
      Collective Influence (CI)             & Influence maximization          & No & Yes      & $O(\lvert V \rvert \log \lvert V \rvert)$                    & \cite{morone_2016} \\
      Belief Propagation-guided Decimation (BPD)        & Message passing-based decycling & No & No      & $O(\lvert E \rvert T) $ & \cite{1mugisha2016identifying}                  \\
      Min-Sum                               & Message passing-based decycling & No & Yes      & $O(\lvert E \rvert T) +O(\lvert V \rvert (\log(\lvert V \rvert) + T )) $ & \cite{Braunstein12368}                  \\
      Generalized Network Dismantling (GND) & Spectral partitioning           & No & Optional & $O(\lvert V \rvert \log^{(2 + \epsilon)}(\lvert V \rvert))$   & \cite{Ren6554}                  \\
      Ensemble GND (EGND)         & Spectral partitioning  & No     & Optional    & $O(e) \cdot O(GND)$            & \cite{10.1007/978-3-030-36687-2_65}     \\
      CoreHD                      & Degree-based decycling & No     & Yes         & $O(\lvert V \rvert)$ (on sparse networks)                 & \cite{Zdeborov__2016}     \\
      Explosive Immunization (EI) & Explosive percolation     & No    & No    &  $O(\lvert V \rvert \log(\lvert V \rvert)) $                        & \cite{PhysRevLett.117.208301}     \\
      GDM                         & Machine Learning       & Yes    & Optional    & $O(h (\lvert V \rvert + \lvert E \rvert))$       & \cite{grassia2021machine}     \\
      CoreGDM                     & Machine Learning       & Yes    & Yes    & $O(h (\lvert V \rvert + \lvert E \rvert))$       & \cite{10.1007/978-3-031-28276-8_8}     \\
      FINDER                      & Machine Learning       & No       & Optional         &  $O(\lvert E \rvert + \lvert V \rvert (1 + \log(\lvert V \rvert))$                         & \cite{fan2020finding}     \\
      \bottomrule
      \end{tabular}
  }
  \caption{Summary of the main algorithms for targeted attacks. For each algorithm, we report the type of attack, whether it is static or dynamic, whether it includes the reinsertion phase of nodes, and the computational complexity. In particular, the computational complexity is reported for sparse networks in terms of the number of nodes $\lvert V \rvert$ and edges $\lvert E \rvert$ of the network. $T$ is a fixed parameter of the Min-Sum algorithm, $\epsilon > 0$, $e$ is the ensemble size, and $h$ is the number of attention heads employed.}
  \label{t:algorithms}
\end{table}

\begin{table}[htbp!]
\begin{adjustbox}{max width=\textwidth}
\begin{tabular}{@{}cl|llllllll|lllll@{}}
\toprule
\multicolumn{1}{l}{} & Network & AD & BC & EI $\sigma_1$ & GDM & GND & FINDER & PR & MS & CI $\ell-2$ & CoreHD & GDM +R & GND +R & MS +R \\
\midrule
\cellcolor[HTML]{BF0041} & arenas-meta & 115.3 & 158.5 & 134.1 & 111.3 & 143.6 & 123.0 & 127.8 & 134.1 & 100.0 & 106.4 & 101.1 & 103.0 & 105.6 \\
\cellcolor[HTML]{BF0041} & dimacs10-celegansneural & 121.3 & 147.3 & 124.4 & 117.7 & 100.0 & 119.4 & 137.0 & 138.3 & 100.0 & 131.4 & 110.9 & 122.2 & 129.9 \\
\cellcolor[HTML]{BF0041} & foodweb-baydry & 106.2 & 118.4 & 111.5 & 108.3 & 113.1 & 105.5 & 130.5 & 113.0 & 100.0 & 109.5 & 105.9 & 106.1 & 107.5 \\
\cellcolor[HTML]{BF0041} & foodweb-baywet & 104.9 & 118.9 & 109.4 & 105.3 & 116.1 & 103.9 & 130.2 & 112.5 & 100.0 & 107.5 & 103.7 & 114.3 & 107.2 \\
\cellcolor[HTML]{BF0041} & maayan-figeys & 105.8 & 161.2 & 163.8 & 105.1 & 109.6 & 104.3 & 133.7 & 136.0 & 100.0 & 129.9 & 102.9 & 105.2 & 129.6 \\
\cellcolor[HTML]{BF0041} & maayan-foodweb & 115.0 & 126.5 & 148.2 & 100.3 & 111.8 & 100.0 & 119.2 & 155.0 & 103.8 & 136.5 & 100.3 & 125.8 & 144.8 \\
\cellcolor[HTML]{BF0041} & maayan-Stelzl & 109.2 & 146.0 & 121.7 & 106.6 & 153.6 & 107.4 & 120.8 & 119.0 & 100.0 & 114.1 & 102.7 & 120.8 & 112.0 \\
\cellcolor[HTML]{BF0041} & maayan-vidal & 119.1 & 142.9 & 117.1 & 115.3 & 128.0 & 112.3 & 125.8 & 131.6 & 100.0 & 110.2 & 103.9 & 118.1 & 112.9 \\
\cellcolor[HTML]{BF0041} & moreno\_propro & 131.5 & 166.9 & 100.0 & 113.8 & 131.8 & 121.4 & 143.5 & 165.2 & 105.1 & 104.9 & 103.2 & 107.6 & 105.9 \\
\multirow{-10}{*}{\cellcolor[HTML]{BF0041}\textbf{\begin{tabular}[c]{@{}c@{}}B\\  I\\  O \end{tabular}}} & {\ul \textit{Biological Result}} & {\ul \textit{114.3}} & {\ul \textit{143.0}} & {\ul \textit{125.6}} & {\ul \textit{109.3}} & {\ul \textit{123.1}} & {\ul \textit{110.8}} & {\ul \textit{129.8}} & {\ul \textit{133.9}} & {\ul \textit{101.0}} & {\ul \textit{116.7}} & {\ul \textit{103.8}} & {\ul \textit{113.7}} & {\ul \textit{117.3}} \\
\midrule
\cellcolor[HTML]{2A6099} & cfinder-google & 147.4 & 559.6 & 346.0 & 148.1 & 237.5 & 167.6 & 168.1 & 1011.1 & 136.0 & 149.5 & 100.0 & 156.4 & 247.2 \\
\cellcolor[HTML]{2A6099} & cit-HepPh & 134.2 & 152.8 & 127.7 & 133.7 & 100.0 & 132.7 & 147.7 & 138.4 & 115.1 & 128.3 & 126.2 & 119.9 & 128.7 \\
\cellcolor[HTML]{2A6099} & citeseer & 119.2 & 180.8 & 115.6 & 116.6 & 119.1 & 116.7 & 139.6 & 129.7 & 100.0 & 109.9 & 108.2 & 106.5 & 110.7 \\
\cellcolor[HTML]{2A6099} & com-amazon & 163.0 & 227.8 & 100.0 & 141.7 & NA & 161.8 & 162.3 & 202.2 & 116.9 & 109.0 & 109.9 & NA & 112.0 \\
\cellcolor[HTML]{2A6099} & com-dblp & 132.6 & 140.3 & 102.6 & 113.1 & 124.0 & 134.1 & 127.0 & 208.8 & 100.0 & 103.2 & 102.8 & 122.5 & 104.5 \\
\cellcolor[HTML]{2A6099} & dblp-cite & 123.6 & 129.5 & 121.6 & 108.9 & 123.3 & 115.9 & 124.9 & 152.2 & 100.0 & 144.0 & 113.4 & 118.1 & 144.2 \\
\cellcolor[HTML]{2A6099} & dimacs10-polblogs & 112.7 & 126.7 & 127.5 & 110.4 & 118.7 & 108.9 & 124.2 & 118.7 & 100.0 & 117.0 & 108.6 & 119.7 & 115.8 \\
\cellcolor[HTML]{2A6099} & econ-wm1 & 112.0 & 135.4 & 130.6 & 102.0 & 133.0 & 100.0 & 133.6 & 109.9 & 131.4 & 108.1 & 101.6 & 111.6 & 108.1 \\
\cellcolor[HTML]{2A6099} & linux & 162.8 & 511.4 & 118.4 & 140.1 & 137.1 & 157.2 & 246.5 & 210.1 & 122.1 & 103.7 & 109.7 & 100.0 & 112.2 \\
\cellcolor[HTML]{2A6099} & p2p-Gnutella06 & 114.9 & 125.3 & 114.9 & 105.8 & 136.0 & 111.6 & 118.1 & 115.1 & 100.0 & 116.5 & 107.4 & 127.4 & 114.7 \\
\cellcolor[HTML]{2A6099} & p2p-Gnutella31 & 112.9 & 134.1 & 116.6 & 103.5 & 138.3 & 109.6 & 114.2 & 113.0 & 100.0 & 114.3 & 106.0 & 125.9 & 112.6 \\
\cellcolor[HTML]{2A6099} & subelj\_jdk & 142.3 & 428.7 & 139.7 & 123.6 & 132.9 & 129.2 & 178.2 & 178.7 & 113.4 & 104.8 & 100.0 & 104.7 & 100.1 \\
\cellcolor[HTML]{2A6099} & subelj\_jung-j & 169.0 & 463.4 & 163.9 & 138.5 & 141.4 & 149.1 & 208.7 & 198.2 & 115.8 & 114.9 & 111.0 & 122.5 & 100.0 \\
\cellcolor[HTML]{2A6099} & web-EPA & 108.6 & 146.3 & 149.9 & 106.2 & 157.6 & 106.0 & 111.5 & 167.9 & 100.0 & 142.2 & 107.4 & 122.8 & 141.1 \\
\cellcolor[HTML]{2A6099} & web-NotreDame & 249.1 & 141.3 & 100.0 & 117.0 & 121.0 & 236.0 & 152.6 & 511.5 & 166.1 & 116.7 & 110.8 & 107.7 & 118.4 \\
\cellcolor[HTML]{2A6099} & web-Stanford & 364.3 & 416.6 & 113.0 & 171.7 & 199.3 & 359.9 & 300.9 & NA & 153.6 & 124.7 & 131.5 & 100.0 & NA \\
\cellcolor[HTML]{2A6099} & web-webbase-2001 & 329.4 & 414.6 & 392.4 & 199.6 & 254.8 & 960.0 & 431.8 & 7192.8 & 2758.4 & 153.0 & 129.1 & 100.0 & 164.9 \\
\cellcolor[HTML]{2A6099} & wordnet-words & 133.8 & 162.4 & 123.7 & 108.4 & 130.5 & 130.2 & 126.5 & 254.1 & 100.0 & 118.9 & 108.3 & 120.1 & 120.3 \\
\multirow{-19}{*}{\cellcolor[HTML]{2A6099}\textbf{\begin{tabular}[c]{@{}c@{}}I\\  N\\  F\\  O\\  R\\  M\\  A\\  T\\  I\\  O\\  N\end{tabular}}} & {\ul \textit{Information Result}} & {\ul \textit{162.9}} & {\ul \textit{255.4}} & {\ul \textit{150.2}} & {\ul \textit{127.2}} & {\ul \textit{147.3}} & {\ul \textit{193.7}} & {\ul \textit{173.1}} & {\ul \textit{647.8}} & {\ul \textit{262.7}} & {\ul \textit{121.0}} & {\ul \textit{110.7}} & {\ul \textit{116.8}} & {\ul \textit{126.8}} \\
\midrule
\cellcolor[HTML]{468A1A} & advogato & 118.1 & 133.0 & 129.3 & 115.9 & 125.1 & 114.1 & 174.0 & 130.6 & 100.0 & 118.4 & 109.9 & 113.0 & 119.0 \\
\cellcolor[HTML]{468A1A} & com-youtube & 117.3 & 173.5 & 159.9 & 120.6 & 113.6 & 116.2 & 134.6 & 148.4 & 100.0 & 125.0 & 111.4 & 108.3 & 124.5 \\
\cellcolor[HTML]{468A1A} & corruption & 161.2 & 170.8 & 242.1 & 102.4 & 101.7 & 148.3 & 151.2 & 886.0 & 283.9 & 142.0 & 100.0 & 151.0 & 143.0 \\
\cellcolor[HTML]{468A1A} & digg-friends & 116.5 & 152.8 & 158.0 & 112.8 & 113.8 & 109.8 & 139.1 & 158.6 & 100.0 & 136.7 & 109.4 & 116.9 & 136.2 \\
\cellcolor[HTML]{468A1A} & douban & 113.7 & 123.5 & 121.4 & 100.7 & 121.6 & 100.0 & 107.6 & 133.5 & 103.1 & 133.8 & 102.1 & 130.2 & 132.5 \\
\cellcolor[HTML]{468A1A} & ego-twitter & 111.9 & 137.0 & 105.8 & 102.7 & 120.0 & 113.8 & 110.7 & 171.8 & 100.0 & 117.5 & 101.5 & 100.8 & 114.8 \\
\cellcolor[HTML]{468A1A} & email-EuAll & 100.9 & 118.8 & 218.9 & 103.1 & 100.0 & 136.3 & 100.7 & 198.0 & 101.9 & 153.1 & 103.1 & 103.1 & 152.0 \\
\cellcolor[HTML]{468A1A} & hyves & 111.4 & 159.4 & 168.9 & 100.4 & 110.6 & 102.5 & 104.1 & 135.2 & 100.0 & 135.2 & 101.6 & 110.9 & 133.5 \\
\cellcolor[HTML]{468A1A} & librec-ciaodvd-trust & 132.0 & 128.9 & 145.2 & 112.2 & 126.9 & 115.8 & 135.2 & 142.0 & 100.0 & 139.6 & 117.0 & 128.3 & 141.6 \\
\cellcolor[HTML]{468A1A} & librec-filmtrust-trust & 135.8 & 183.5 & 130.2 & 115.4 & 125.7 & 131.0 & 152.1 & 194.7 & 100.0 & 123.3 & 103.5 & 110.2 & 113.9 \\
\cellcolor[HTML]{468A1A} & loc-brightkite & 119.2 & 147.5 & 118.1 & 120.8 & 121.1 & 116.3 & 126.0 & 129.0 & 100.0 & 111.3 & 108.2 & 120.5 & 111.6 \\
\cellcolor[HTML]{468A1A} & loc-gowalla & 126.8 & 177.1 & 120.8 & 126.8 & 130.9 & 125.6 & 136.6 & 133.6 & 100.0 & 114.8 & 113.7 & 116.5 & 115.3 \\
\cellcolor[HTML]{468A1A} & moreno\_crime\_projected & 231.2 & 218.4 & 168.3 & 120.9 & 127.9 & 206.5 & 190.6 & 1180.7 & 128.0 & 121.3 & 100.0 & 107.3 & 125.9 \\
\cellcolor[HTML]{468A1A} & moreno\_train & 107.1 & 134.7 & 124.0 & 100.0 & 104.9 & 108.8 & 149.5 & 176.9 & 160.4 & 115.6 & 100.0 & 109.7 & 120.3 \\
\cellcolor[HTML]{468A1A} & munmun\_digg\_reply\_LCC & 108.9 & 129.9 & 120.9 & 110.5 & 128.5 & 107.5 & 117.7 & 109.3 & 100.0 & 109.4 & 105.7 & 114.9 & 108.8 \\
\cellcolor[HTML]{468A1A} & munmun\_twitter\_social & 111.7 & 113.9 & 137.7 & 108.6 & 114.2 & 113.4 & 100.0 & 152.6 & 108.4 & 149.1 & 108.8 & 122.1 & 150.4 \\
\cellcolor[HTML]{468A1A} & opsahl-ucsocial & 111.8 & 122.9 & 132.6 & 111.9 & 136.6 & 108.9 & 118.5 & 121.7 & 100.0 & 118.3 & 108.5 & 118.8 & 118.6 \\
\cellcolor[HTML]{468A1A} & pajek-erdos & 109.1 & 112.6 & 126.5 & 105.6 & 118.5 & 107.8 & 109.0 & 129.6 & 100.0 & 123.2 & 103.7 & 112.9 & 120.3 \\
\cellcolor[HTML]{468A1A} & petster-cat-household & 110.6 & 197.9 & 148.3 & 114.7 & 114.4 & 108.3 & 132.8 & 197.1 & 100.0 & 179.6 & 110.5 & 111.6 & 177.6 \\
\cellcolor[HTML]{468A1A} & petster-catdog-household & 108.8 & 137.6 & 123.5 & 107.5 & 108.2 & 106.7 & 125.9 & 177.1 & 100.0 & 155.6 & 101.8 & 105.4 & 154.2 \\
\cellcolor[HTML]{468A1A} & petster-hamster & 145.1 & 146.4 & 122.8 & 118.2 & 109.4 & 146.0 & 159.7 & 197.1 & 100.0 & 113.8 & 108.2 & 110.3 & 114.1 \\
\cellcolor[HTML]{468A1A} & slashdot-threads & 108.2 & 124.2 & 133.2 & 108.8 & 108.9 & 106.4 & 113.7 & 128.0 & 100.0 & 125.8 & 104.5 & 104.2 & 125.1 \\
\cellcolor[HTML]{468A1A} & slashdot-zoo & 108.7 & 141.0 & 137.4 & 113.6 & 112.7 & 106.7 & 117.4 & 127.7 & 100.0 & 121.5 & 108.1 & 111.1 & 120.3 \\
\cellcolor[HTML]{468A1A} & soc-Epinions1 & 116.8 & 139.5 & 134.1 & 115.8 & 115.5 & 112.5 & 120.8 & 136.2 & 100.0 & 120.4 & 110.0 & 111.4 & 120.5 \\
\cellcolor[HTML]{468A1A} & twitter\_LCC & 136.6 & 241.0 & 116.2 & 129.1 & 125.1 & 132.4 & 161.1 & 132.0 & 100.0 & 113.2 & 111.3 & 108.7 & 110.9 \\
\multirow{-26}{*}{\cellcolor[HTML]{468A1A}\textbf{\begin{tabular}[c]{@{}c@{}}S\\  O\\  C\\  I\\  A\\  L\end{tabular}}} & {\ul \textit{Social Result}} & {\ul \textit{123.2}} & {\ul \textit{150.6}} & {\ul \textit{141.8}} & {\ul \textit{112.0}} & {\ul \textit{117.4}} & {\ul \textit{120.1}} & {\ul \textit{131.5}} & {\ul \textit{221.1}} & {\ul \textit{111.4}} & {\ul \textit{128.7}} & {\ul \textit{106.5}} & {\ul \textit{114.3}} & {\ul \textit{128.2}} \\
\midrule
\cellcolor[HTML]{FFFF00} & ARK201012\_LCC & 118.0 & 140.8 & 146.6 & 114.2 & 113.8 & 119.4 & 117.7 & 149.5 & 106.1 & 105.7 & 107.9 & 100.0 & 109.4 \\
\cellcolor[HTML]{FFFF00} & dimacs9-COL & 2175.8 & 1218.2 & 100.0 & 954.0 & 394.5 & 1944.9 & 1889.4 & 2000.9 & 1526.1 & 202.9 & 168.1 & 217.0 & 173.8 \\
\cellcolor[HTML]{FFFF00} & dimacs9-NY & 1515.0 & 473.0 & 123.8 & 365.9 & 187.7 & 1326.4 & 1343.4 & 1513.4 & 777.4 & 143.5 & 155.2 & 100.0 & 152.3 \\
\cellcolor[HTML]{FFFF00} & eu-powergrid & 215.5 & 270.9 & 114.6 & 155.2 & 117.8 & 230.0 & 279.5 & 451.6 & 166.8 & 129.5 & 100.0 & 103.3 & 144.1 \\
\cellcolor[HTML]{FFFF00} & gridkit-eupowergrid & 373.3 & 482.4 & 100.0 & 267.8 & 175.6 & 362.8 & 362.9 & 702.8 & 276.7 & 102.4 & 121.2 & 101.9 & 110.3 \\
\cellcolor[HTML]{FFFF00} & gridkit-north\_america & 474.7 & 568.8 & 100.0 & 361.6 & 214.6 & 521.8 & 508.5 & 700.8 & 341.1 & 136.0 & 167.6 & 110.4 & 122.3 \\
\cellcolor[HTML]{FFFF00} & inf-USAir97 & 140.5 & 138.5 & 158.3 & 107.7 & 121.1 & 127.9 & 126.1 & 176.6 & 100.0 & 111.6 & 107.8 & 126.3 & 115.9 \\
\cellcolor[HTML]{FFFF00} & internet-topology & 116.9 & 145.0 & 134.4 & 118.0 & 112.8 & 117.1 & 128.8 & 163.5 & 102.5 & 118.2 & 111.9 & 100.0 & 120.1 \\
\cellcolor[HTML]{FFFF00} & london\_transport\_aggr & 143.1 & 169.4 & 124.1 & 116.8 & 103.6 & 142.9 & 148.6 & 137.7 & 115.8 & 121.3 & 100.0 & 104.7 & 117.2 \\
\cellcolor[HTML]{FFFF00} & opsahl-openflights & 150.7 & 146.4 & 130.2 & 118.4 & 119.9 & 134.3 & 145.9 & 186.3 & 112.5 & 121.5 & 100.0 & 109.0 & 131.8 \\
\cellcolor[HTML]{FFFF00} & opsahl-powergrid & 402.7 & 498.5 & 100.2 & 271.0 & 100.0 & 402.0 & 469.9 & 445.2 & 272.9 & 139.4 & 116.9 & 114.1 & 142.3 \\
\cellcolor[HTML]{FFFF00} & oregon2\_010526 & 135.3 & 142.5 & 162.9 & 124.3 & 132.7 & 124.1 & 126.3 & 201.3 & 105.4 & 140.5 & 111.8 & 100.0 & 140.1 \\
\cellcolor[HTML]{FFFF00} & power-eris1176 & 392.9 & 496.9 & 198.3 & 115.5 & 229.9 & 392.8 & 292.8 & 730.6 & 186.3 & 186.3 & 100.0 & 178.8 & 182.3 \\
\cellcolor[HTML]{FFFF00} & roads-california & 869.5 & 3978.3 & 105.5 & 415.7 & 351.8 & 524.4 & 564.1 & 518.3 & 445.7 & 106.7 & 163.6 & 198.5 & 100.0 \\
\cellcolor[HTML]{FFFF00} & roads-northamerica & 1117.8 & 8119.5 & 100.0 & 555.9 & 1076.6 & 972.9 & 809.8 & NA & 706.4 & 171.6 & 170.6 & 364.3 & NA \\
\cellcolor[HTML]{FFFF00} & roads-sanfrancisco & 1190.4 & 333.1 & 100.0 & 593.7 & 193.4 & 1138.1 & 1669.5 & 1035.1 & 527.9 & 162.0 & 156.6 & 138.3 & 194.9 \\
\cellcolor[HTML]{FFFF00} & route-views & 124.2 & 137.0 & 162.5 & 122.0 & 121.2 & 126.1 & 126.3 & 160.4 & 112.2 & 113.4 & 114.7 & 100.0 & 116.2 \\
\cellcolor[HTML]{FFFF00} & tech-RL-caida & 153.8 & 219.9 & 100.0 & 136.2 & 142.8 & 148.1 & 154.6 & 200.5 & 115.1 & 109.2 & 119.2 & 128.4 & 112.8 \\
\multirow{-19}{*}{\cellcolor[HTML]{FFFF00}\textbf{\begin{tabular}[c]{@{}c@{}}T\\  E\\  C\\  H\\  N\\  O\\  L\\  O\\  G\\  I\\  C\\  A\\  L\end{tabular}}} & {\ul \textit{Technological Result}} & {\ul \textit{545.0}} & {\ul \textit{982.2}} & {\ul \textit{125.6}} & {\ul \textit{278.6}} & {\ul \textit{222.8}} & {\ul \textit{492.0}} & {\ul \textit{514.7}} & {\ul \textit{557.3}} & {\ul \textit{338.7}} & {\ul \textit{134.5}} & {\ul \textit{127.4}} & {\ul \textit{138.6}} & {\ul \textit{134.5}} \\
\midrule
\multicolumn{1}{l}{} & {\ul \textit{Grand Average}} & {\ul \textit{240.5}} & {\ul \textit{390.0}} & {\ul \textit{137.6}} & {\ul \textit{158.3}} & {\ul \textit{153.0}} & {\ul \textit{233.2}} & {\ul \textit{240.4}} & {\ul \textit{399.7}} & {\ul \textit{207.2}} & {\ul \textit{126.6}} & {\ul \textit{112.6}} & {\ul \textit{121.2}} & {\ul \textit{127.9}} \\
\bottomrule
\end{tabular}

\end{adjustbox}
\caption{Per-method Area Under the Curve (AUC) of the dismantling of real-world networks, grouped per category. The dismantling target for each method is $10\%$ of the network size. We compute the AUC value by integrating the $\lvert \mathrm{LCC}(x) \rvert /\lvert V \rvert$ --- i.e., the Largest Connected Component's size as a function of the removed nodes --- values using Simpson’s rule. For sake of readability, the AUC values of each network are expressed as the percentage of the value obtained by the best performing method on that network. That is, the lower the AUC, the better. Column ``AD'' stands for Adaptive Degree, ``BC'' for Betweenness Centrality, ``CI'' for Collective Influence, ``EI'' for Explosive Immunization, ``GND'' for Generalized Network Dismantling, ``MS'' for MinSum, and ``PR'' for PageRank. ``+R'' means that the reinsertion phase is performed. We note that CoreHD and CI are compared to other +R algorithms as they include the reinsertion phase. Some results are missing (``-'') as the algorithm did not return correctly.}

\label{t:real_world_results}
\end{table}

Except from the ad-hoc algorithms shown in Table~\ref{t:algorithms} one could employ direct optimization algorithms (like Simulated Annealing, Tabu search or genetic algorithms) to find the optimal dismantling set $q_c$.
However, due to to large dimension of the search space they do not scale well and, thus, are often unable to provide good solutions. 
As an example, as shown in~\cite{Zdeborov__2016}, the MinSum algorithm outperforms Simulated Annealing.

So far we have approached robustness and resilience from the static viewpoint of percolation models. However, it is desirable to address the scenario in which small malfunctions, located in the network either randomly or in a targeted fashion, can spread and trigger global network-wide effects, the so-called cascades. We investigate them in the next section.

\section{Cascading failures}\label{sec:cascading_failures}

\begin{figure}[h]
    \centering
    \includegraphics[width=0.95\linewidth]{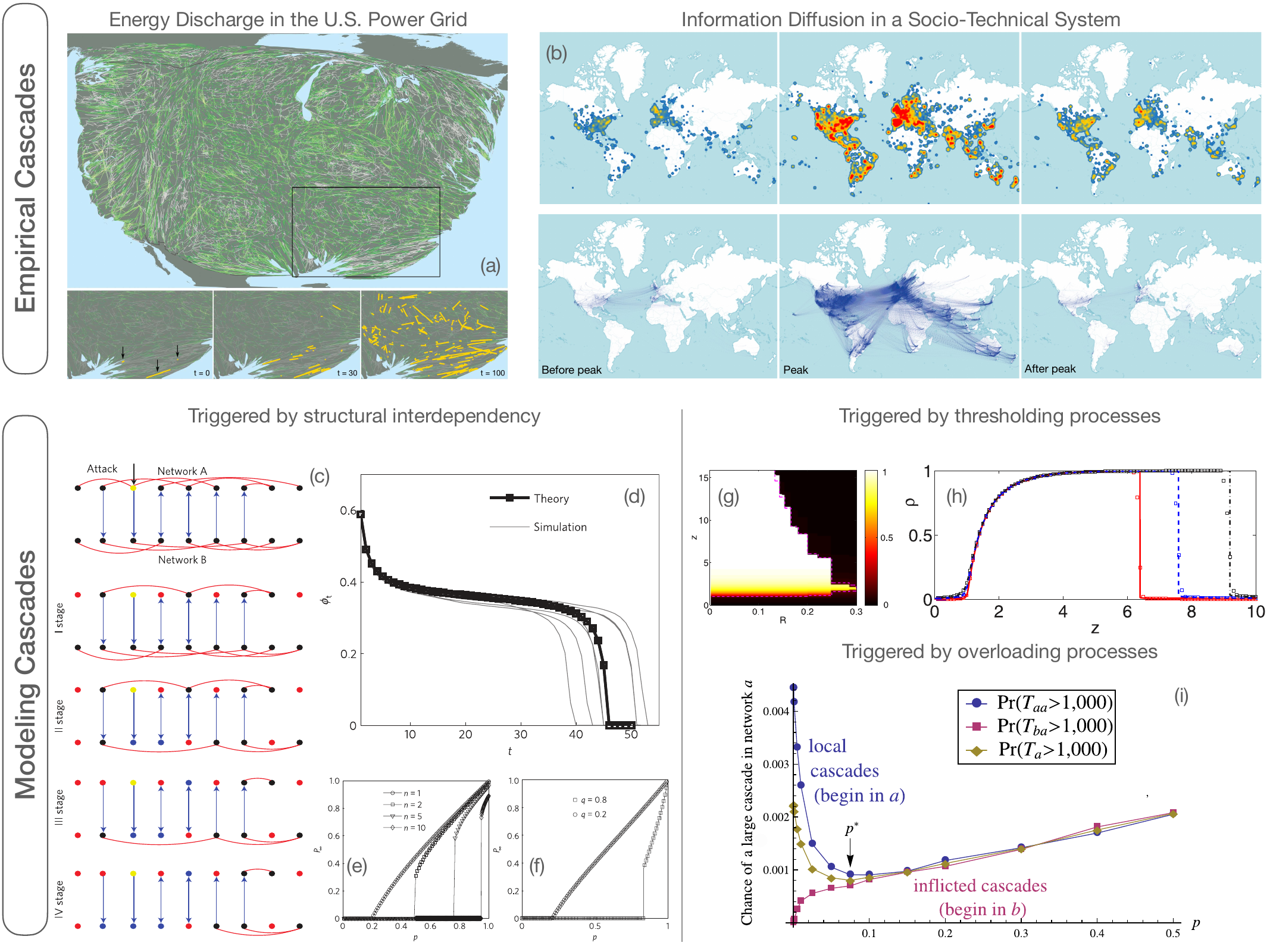}
    \caption{\footnotesize{\textbf{Evolving failures on networks.} (a) Risk assessment to cascade spreading in the U.S.-South Canada power grid (from~\cite{yang2017small}). Cartogram with the state of power lines after the simulations of cascade spreading: lines that never underwent outage (green) vs. affected power lines (gray). In the bottom row, snapshots of the evolution of the damaged lines (in yellow) after a cascade triggered by 3 failures at a rescaled time $t=0$. Notice the nonlocal behavior, characteristic of overload cascading failures. (b) Information cascade in Twitter, showing the density of tweets (top row) about the discovery of the Higgs boson before, during, and after the Nobel announcement, as well as the corresponding network of re-shared messages (bottom row). Note the global reaches of the cascade at the peak, compared to the localized behavior of the information spreading before and after the cascading event (from~\cite{de2013anatomy}). (c) Sketch of the evolution of a failure sustained by dependency relations in a toy system with two coupled layers. Directed vertical arrows represent dependency relations. Functional nodes are represented in black; the node that initially fails is indicated in yellow; red and blue nodes indicate removed units because either they do not belong to the largest cluster (red) or they depend on failed nodes in the other network (blue). (d) Size of the giant component as a function of the cascade step. Grey lines correspond to individual realizations of the dynamics, the markers indicated their average. The black line is the theoretical prediction. (e) Stationary size of the giant component as a function of the fraction of initially removed nodes. The different markers correspond to different number of coupled layers, and solid lines show the theoretical predictions. Same for panel (f), but the different markers/lines correspond to the fraction of interdependent nodes in a bilayer network. Panels (c-f) are reproduced from~\cite{gao2012networks}. (g) Size of the giant component at the cascade stop for the threshold model introduced by~\cite{watts2002simple} and discussed in the main text, as a function of the threshold $R$ and the mean degree $\langle k \rangle$ of the underlying Erdos-Renyi network. The dashed line indicates the analytical prediction for the cascade emergence. (h) Size of the giant component at the cascade stop for fixed threshold $R=0.18$, as a function of the mean degree. Different markers/lines, from left to right, correspond to initial seeds of $10^{-3}$, $5 \times 10^{-3}$ and $10^{-2}$. Figures (g) and (h) are reproduced from~\cite{gleeson2007seed}. (i) Large load-shedding cascades can be mitigated at a nontrivial intermediate level of interconnectivity $p$ between two networks (golden curve, diamond markers). However, too many or too few interconnections induce larger cascades and become detrimental for the robustness. $T_{aa}$ ($T_{ba}$) is the cascade size unfolded on network $a$ for cascades that started at network $a$ ($b$). $T_a$ is the size of cascades in network $a$ without distinguishing where the cascade begins. Network $a$ and $b$ have $2000$ nodes. Figure from~\cite{brummitt2012suppressing}.
    }
    }
    \label{fig:FigCasc}
\end{figure}

One of the most devastating processes that can unfold in a network is a cascade of failures. The errors and attacks suffered by empirical systems usually are dynamical in nature  and, even if they originate in a small, localized part of the system, they may  spread further until the entire system catastrophically collapses. Examples range from line outages in the power  grid (see Fig.~\ref{fig:FigCasc}(a)), to mass extinctions in ecological areas, to fake news or rumor spreading in online social platforms (see Fig.~\ref{fig:FigCasc}(b)). It is of crucial importance to understand the mechanisms under which such malfunctions are able to propagate, and how the network is affected in the meanwhile. In spite of the differences with the static percolation framework, some metrics borrowed from percolation are used to quantify the robustness and resilience to cascades, such as the size of the giant component of the non-failed network once the cascade stops, or the critical point at which the surviving giant component dismantles. In this section, we review different approaches to cascading failures by using stylized models that seek to provide general insights into malfunction (or other) spreading. We do so from a statistical physics standpoint, namely, by disregarding the particular microscopic, domain-oriented aspects of particular systems that may go beyond this review. Instead, we advocate for focusing attention on the large-scale patterns and the collective, emergent behaviors that are common to a variety of seemingly different networked systems. 

One large family of models are those in which failures spread through interdependencies. This is a central concept in systems theory, and refers to the dependency relations that exist between different micro and mesoscopic parts of a macroscopic system and that sustain the overall proper functionality. Examples pervade many branches of sciences across scales, from biochemistry to man-made planetary engineering systems~\cite{strogatz2022fifty}. Two equivalent theoretical frameworks, i.e., multilayer networks~\cite{kivela2014multilayer, boccaletti2014structure, bianconi2018multilayer, artime2022multilayer} and networks of networks~\cite{kenett2015networks, gao2022introduction}, have been successfully employed to frame the modelling of such cascades. In them, each layer (or network) is associated with a subsystem and dependencies are encoded in the interlayer (or internetwork) links. At odds with static percolation, in this case one deals with snapshots of a network that is progressively degraded, whose dismantling is dictated by dependency-related rules (see Fig.~\ref{fig:FigCasc}(c) for details).

In 2003, the Italian grid suffered a power outage due to a damaged power station that led to the failure of some Internet communication networks, which, in turn, created a feed-forward loop of failures among both systems. Inspired by this, Buldyrev and coauthors developed a model of casacading failures mediated by interdependencies in a bilayer network~\cite{buldyrev2010catastrophic}. Using their model, percolation quantities, such as the size of the giant component, can be analytically tracked at each snapshot, finding excellent agreement with simulations (see Fig.~\ref{fig:FigCasc}(d)). The theory is flexible and admits generalizations to an arbitrary number of coupled subsystems~\cite{gao2011robustness, gao2012networks}, partial and asymmetric interdependency relations~\cite{parshani2010interdependent, schneider2013towards}, etc. For instance, if layer $i=1,\ldots,n$ suffers a failure or an attack that leaves a fraction $\phi_i$ of functional nodes, and $q_{ji}$ indicates the fraction of nodes in layer $i$ that directly depend on nodes of layer $j$, then the stationary value of the giant component in layer $i$, $S_{i}^{\text{\,st}}$, can be obtained by solving the system of equations~\cite{gao2012networks}
\begin{align}
S_{i}^{\text{\,st}} & = x_i g_i(x_i), \label{eq:casc1} \\
x_i & = \phi_i \prod_{j=1}^{K} \left[ q_{ji} y_{ji} g_j(x_j) - q_{ji}+ 1 \right], \label{eq:casc2}\\
y_{ij} & = \frac{x_i}{q_{ji} y_{ji}g_j(x_j) - q_{ji}+1}, \label{eq:casc3}
\end{align}
where $g_i(z) \equiv 1 - G_i(z f_i(z) + 1 - z)$ and $f_i(z) = H_i(z f_i(z) + 1 -z)$. The degree and excess degree generating functions of subsystem $i$, that has degree distribution $p_k^i$, are respectively given by $G_i(x) \equiv \sum_k p_k^i x^k$ and $H_i(x)= G_i'(x) / G_i'(1)$. $K$ is the number of layers connected to $i$ by interdepedence links. An analytical treatment of cascades driven by link failures is also possible~\cite{chen2020robustness}, yet this case has been scarcely studied in comparison with the node failure propagation mechanisms.

Eqs.~\eqref{eq:casc1}-\eqref{eq:casc3} predict phenomenologically rich scenarios, which, however, have dramatic consequences for the robustness of interdependent systems. The most striking one is related to the emergence of a discontinuous transition in the giant connected component at the cascade stop. That translates into abrupt system collapses, that are more difficult to anticipate as the number of fully coupled subsystems increases (see Fig.~\ref{fig:FigCasc}(e)). This can be attenuated by decreasing the level of interdependency among layers, recovering the continuous dismantling if it is reduced enough (see Fig.~\ref{fig:FigCasc}(f)). Most importantly, the analysis of Eqs.~\eqref{eq:casc1}-\eqref{eq:casc3} offer hints on how to improve the robustness of interdependent coupled systems. Three main methods have been identified: (i) increase the fraction of autonomous nodes~\cite{parshani2010interdependent}, specially those with high degree~\cite{schneider2013towards}; (ii) design dependency relations between nodes of similar degree~\cite{parshani2011inter, buldyrev2011interdependent}; (iii) devote special efforts to protect high-degree nodes against failures and attacks~\cite{schneider2013towards}. In the context of brain networks, it has been shown that robustness to interdependency-mediated cascades can be improved if topological correlations are taken into account~\cite{reis2014avoiding}. We refer the reader to the recent reviews~\cite{shekhtman2016recent, valdez2020cascading} and references therein for a more in-depth discussion on this type of cascades.

Another family of models has its origin in theories of collective behavior, the so-called threshold models. In these,  we assume that the influence exerted by the neighbors of a node on its probability of changing state, i.e., going from functional to failed, is not linear in the number $n$ of failed neighbors, and in fact only takes effect above some threshold value of $n$~\cite{schelling1973hockey, granovetter1978threshold, easley2010networks}. This class of models has frequently been framed in the context of social psychology, e.g., unraveling under which circumstances a social agent will decide to spread rumours, fads, fake news, or to adopt behavior,  etc. This is of course relevant to the context of network robustness and resilience at a societal level because it is well known that the information spreading here can represent a serious threat to public health systems, polarize debates and undermine public trust in democratic institutions~\cite{gallotti2020assessing,watts2021measuring}, but spreading can also have positive impacts such as diffusing innovations~\cite{valente2005network,easley2010networks}. In any case, these models can be easily framed as cascades occurring in other areas of interest.

The piece-wise, non-continuous influence of neighbors upon a node in a threshold model can be implemented in several ways. One option, considered by Watts in his seminal paper~\cite{watts2002simple}, is to assume that a node will fail if a fraction, larger than a certain threshold, of its neighbors has also failed. This simple model leads to a rich phenomenology, namely, the emergence of global cascades that is bounded in the parameter space by two transitions, one continuous and one discontinuous, see Figs.~\ref{fig:FigCasc}(g) and (h). Indeed, assuming that the initial number of failed nodes is small compared to the system size (see~\cite{gleeson2007seed, liu2012cascading} for the role played by the initial cascade seeding) and that the threshold is the same for all nodes (the same phenomenology is observed for heterogeneous thresholds~\cite{watts2002simple}), then global cascades are very rare when the threshold is large. When it is reduced, though, the network topology starts to play a role. If the mean degree is too small, there are very few nodes that can spread failures and they are isolated from each other, hence global cascades cannot develop. If the mean degree is large enough, the cascade cannot either evolve because the large number of functional neighbors will stabilize the nodes to be always below the threshold. The lower critical mean degree is barely dependent on the threshold but the upper one significantly depends on it. It is between these two points that global cascades are easily triggered. The condition for the emergence of global cascades and estimates for their size have been successfully generalized to networks with topological correlations~\cite{centola2007cascade, gleeson2008cascades, dodds2009analysis, hackett2011cascades, snyder2022degree}, as well as temporal~\cite{karimi2013threshold, backlund2014effects} and multiplex networks~\cite{brummitt2012multiplexity}. Similarly, more complex failing conditions have been employed, such as combining relative and absolute thresholds~\cite{yu2016system}, setting absolute thresholds only~\cite{galstyan2007cascading}, or including a memory of past exposures to failures~\cite{dodds2004universal}. 

The final set of models we discuss in this section are overload models. In these, nodes (or links) sustain a load, which can be, for example, electrical current in the power grid, network packets in the Internet, passengers in public transportation systems, or the amount of a certain product (wheat, rice, maize, etc.) in an international trade network. The nodes remain functional as long as their load is kept below a threshold, the so-called capacity, which, a priori, can be modified, e.g., by human interventions. After a failure or an attack, load is redistributed according to some rules, which will depend on the type of system one aims to model, causing a potential chain of overloads and new redistribution events across the network. The cascade finishes when all nodes restore their load below their capacity.

One of the most paradigmatic redistribution mechanisms is that of the sand-pile models of self-organized criticality (SOC)~\cite{bak1987self, bak1988self}. These are characterized by a slow external driving, e.g., by introduction of load in the system, and very fast relaxations, where the cascade occurs, e.g., with overloaded nodes shedding load to their neighbors. The relaxation is considered to be so fast that no load is added during its evolution, which usually is a good approximation to empirical processes because the unfolding of a cascade is the fastest time scale. Since networks have nodes with a variable number of neighbors, a natural choice is to set the node capacity equal to the degree~\cite{bonabeau1995sandpile}, although other choices are possible~\cite{lise2002nonconservative, goh2003sandpile, lee2004sandpile}. Moreover, to avoid a network becoming saturated with load, one needs to dissipate (remove) load with a certain probability when it is shed. Load shedding cascade models have been analyzed in the context of interdependent power grids~\cite{brummitt2012suppressing}, where an optimal level of coupling among networks to reduce the impact of the cascades has been found, see Fig.~\ref{fig:FigCasc}(i). Interesting phenomenology has been also reported, such as dragon king events (cascades of larger cascades) when the SOC rules are coupled to the Kuramoto model~\cite{mikaberidze2022sandpile}.

Other load redistribution mechanisms are, of course, also possible. Motter and Lai proposed to further extend the interplay between structure and dynamics by associating load to betweenness~\cite{motter2002cascade}, since there is a positive correlation between shortest paths and traffic/route assignment. In their model each node
has a capacity proportional to its initial betweenness
and when a node fails there is a global load redistribution, possibly causing the overload of new nodes that are not necessarily in the vicinity of the previous failure. This nonlocality can be easily visualized in spatially embedded networks~\cite{zhao2016spatio}. A full, explicit characterization of nonlocality is still an analytical challenge, but some efforts have been made from different perspectives to better understand its complexities~\cite{daqing2014spatial, hines2016cascading, schafer2018dynamically, valente2022non}. Motter and Lai found numerically that networks with heteregoneous degree distributions, thus with heteregoneous betweeness distribution~\cite{barthelemy2004betweenness}, are incredibly weak to targeted attacks if the node capacity is not large enough. In fact, the network could crash even if only one node, chosen among those with either largest betweenness or degree, is taken down as initial stressor. On the other hand, picking uniformly at random a single node as initial perturbation is not an efficient strategy to dismantle a network, even if the capacity of the nodes is low, and independently of the homogeneous or heterogeneous connectivity of the network. However, it has been analytically shown that for a large enough set of randomly chosen nodes, the network can suffer a discontinuous dismantling transition
~\cite{kornbluth2018network}. Numerical evidence for a discontinuous transition has been further reported in both mono and multiplex networks~\cite{artime2020abrupt}. Overload models have also been studied in the context of link failures and interventions, see, e.g.,~\cite{moreno2003critical, lai2004attacks, wang2008universal, cao2013improving, pahwa2014abruptness}.

\section{Preventing and Reacting to Network Collapse}\label{sec:response}

A network collapse can be prevented by employing design principles during the formative stages of a network that make it resilient to random failures and targeted attacks~\cite{paul2004optimization, latora2005vulnerability, Reis2014avoid, carchiolo2019network,10.1007/978-3-030-02738-4_23}. However, such design principles are not always employed, or worse, can be outsmarted by advanced attack strategies that were not foreseeable during network formation. In such cases, early-warning indicators of an impending network collapse~\cite{chen2012detecting,squartini2013early, suweis2014early, dakos2014critical, kuehn2015early,bauch2016early, grassia2021machine} as well as repairs and adaptive responses to an already evolving collapse~\cite{Majdandzic2016optimal,Sun2020optimal,sanhedrai2022reviving} become the best line of defense. Either by favoring spontaneous recovery~\cite{majdandzic2014spontaneous,lin2020non} or by inducing it through microscopic interventions~\cite{Majdandzic2016optimal,zhou2018network,Sun2020optimal,sanhedrai2022reviving}, with cost-efficient network-based procedures~\cite{pan2018resilience,smith2019competitive}, has the potential to prevent systemic collapse.

The earliest attempts of network design can be traced back to augmentation problems in graphs. Already in 1976 Eswaran and Tarjan determined the minimum number of edges necessary to make a directed graph strongly connected and to make an undirected graph bridge-connected or biconnected. Achieving this would mean that if any one vertex were to be removed, the graph would still remain connected. More contemporary network design strategies were proposed by Paul et al.~\cite{paul2004optimization} and by Latora and Marchiori~\cite{latora2005vulnerability}. In particular, Paul et al. have shown that for scale-free networks and for networks with two-peak degree distributions an optimal robustness is designed so that all but one of the nodes have the same degree, which should be close to the average number of links per node, while the one remaining node has a very large degree $k \propto N^{2/3}$, where $N$ is the number of nodes that form the network. Latora and Marchiori, on the other hand, have proposed an approach based on the relative drop in network performance and its minimization given a set of improvements consisting of either adding or removing specific links. More precisely, if $D$ denotes the set of possible damages that can be inflicted upon a network $S$, and if $\ell(S,d)$ is a map that yields the new network after the damage $d \in D$, then we can measure the importance of the damage $d$ by the relative drop in the performance $\Delta\Phi^{-}/\Phi$, where $\Delta\Phi^{-}=\Phi(S)-\Phi[\ell(S,d)] \geq 0$. Moreover, we can define the critical damage $d^* \in D$ as the damage that minimizes $\Phi[\ell(S,d)]$. The vulnerability $V$ of $S$ due to $D$ can then be defined as~\cite{latora2005vulnerability}
\begin{equation}
V(S,D)=\frac{\Phi(S)-W(S,D)}{\Phi(S)}\, ,
\label{resp:v}
\end{equation}
where $W(S,D)=\Phi[\ell(S,d^*)]$ is the worst performance of $S$ under the class of damages $D$. Analogously, we can define a set of improvements $i \in I$ to be made on the network $S$ via the map $\aleph(S,i)$, such that the best improvability of the network subject to the critical improvement $i^*$ is
\begin{equation}
M(S,I)=\frac{B(S,I)-\Phi(S)}{\Phi(S)}\, ,
\label{resp:m}
\end{equation}
where $B(S,I)=\Phi[\aleph(S,i^*)]$ is the best performance of $S$ under the class of improvements $I$. And as for the damages, we can measure the importance of $i$ by the relative increase in the performance $\Delta\Phi^{+}/\Phi$, where $\Delta\Phi^{+}=\Phi[\aleph(S,i)]-\Phi(S)$~\cite{latora2005vulnerability}. Nevertheless, it has been recently shown that there is a fundamental limit to what can be optimized: in fact, network robustness and its performance are competitive features hard to simultaneously maximize~\cite{pasqualetti2020fragility}.

In 2014, Reis et al.~\cite{Reis2014avoid} have further extended the concept of network design to interdependent networks~\cite{di2016recovery,shekhtman2016recent}, showing that the stability of a system of networks relies on the relation between the internal structure of a network and its pattern of connections to other networks. More precisely, for an interdependent network to be considered robust by design, the interconnections should be provided by network hubs, while the connections between networks should be moderately convergent. In this case, a system of networks can be considered stable and robust to failure, which has also been tested and proven correct experimentally using functional brain networks in task and resting states~\cite{Reis2014avoid}.

But despite best design practices, a network collapse oftentimes remains unavoidable. Early detection is in such cases crucial. One of the first attempts in this regard is due to Squartini et al.~\cite{squartini2013early}, who showed that many topological properties of the Dutch interbank network displayed an abrupt change just prior to the 2008 global financial crisis. Suweis and D'Odorico~\cite{suweis2014early} went a step further for socio-ecological networks, proposing an early-warning indicator based on the maximum element of the covariance matrix of the
network, and showing it is an effective leading mark of network instability. Of note, a similar concept has been outlined also for coupled human-environment systems~\cite{bauch2016early}, although not based on the network formalism but on the tipping point theory. Dakos and Bascompte~\cite{dakos2014critical} then combined the theory on tipping points with patterns of network structure to develop critical slowing-down indicators as early-warning signals for detecting the proximity to a potential tipping point in structurally complex ecological communities. Based on 79 empirical mutualistic networks, their research demonstrated success in identifying specialist species as likely the best-indicator for monitoring the proximity of a community to collapse.

\begin{figure}[h]
    \centering
    \includegraphics[width=0.98\linewidth]{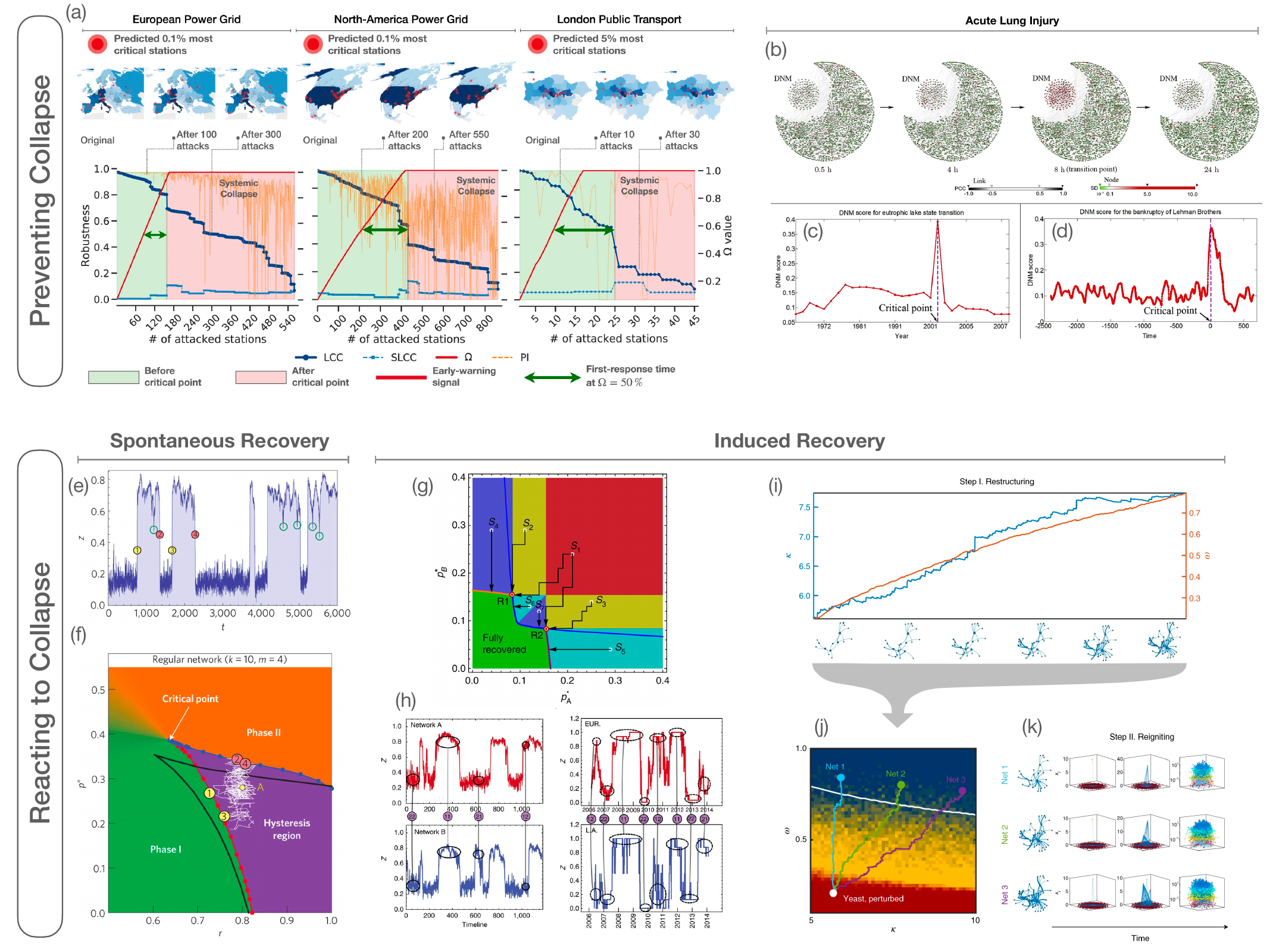}
\caption{\textbf{Preventing and reacting to network collapse.} (a) Early-warning signal measured by $\Omega$ (see text) for three distinct network infrastructures (shown in the top row) repeatedly attacked using a degree-based protocol, and size of the largest and second-largest connected components (bottom row) (from~\cite{grassia2021machine}). (b-d) Dynamical changes in a molecular network from lung tissue of mouse, with a critical transition at 8 hours (top) and node color encoding the fluctuation strength in gene expressions which correlates with a dynamical network marker (DNM) introduced in~\cite{liu2015identifying}; DNM has been used for capturing early-warning signals in other systems, such as eutrophic lake states (left) and daily prices of interest-rate swaps in the USD and EUR currency (right). Figure readapted from~\cite{liu2015identifying}. (e-f) Regular network ($k=10$, $m=4$, $N=100$) with a fraction $z$ of active nodes flipping between two collective modes over time (top) and trajectory of the system in the phase diagram, from $t=0$ to the moment of the first transition (bottom), for the marked states. (g-h) An optimal repair strategy~\cite{Majdandzic2016optimal} for a system with two networks, characterized by a fraction $p_{A}^{\star}$ and $p_{B}^{\star}$ of internally failed nodes. Given the initial state $S_i$ of a collapsed system, the repairing corresponds to minimize the distance between $S_i$ and the nearest border of the green region, where the systems goes back to a fully functional state. Arrows indicate the followed trajectories, while $R_{1}$ and $R_{2}$ are triple points. Collective states, identified by marked numbers, for two synthetic (left) and empirical (right) coupled networks are show, in the bottom. (i-k) Restructuring and reigniting: a 2-steps procedure that drives a perturbed yeast protein interaction network from a collapsed phase (red) into the recoverable phase (blue). Figure from~\cite{sanhedrai2022reviving}.}
    \label{fig:FigRecovery}
\end{figure}

Most recently in this line of research, Grassia et al.~\cite{grassia2021machine} have shown that a machine trained to dismantle relatively small systems is able to identify higher-order topological
patterns and thus effectively disintegrate also large-scale social, infrastructural, and technological networks. And do so more efficiently than human-based heuristics. Remarkably, it was also shown that this can be reversed engineered -- as the machine could assesses the probability of future attacks to disintegrate the system -- for developing early-warning signals of network collapse. More precisely, if $S_o$ is the set of virtually removed nodes that cause the percolation of the network and 
\begin{equation}
	\Omega_{m} = \sum\limits_{n\in S_o}^{}{p_n}\, ,
\end{equation}
then the value of the early warning $\Omega$ for the network after the removal of a generic set $S$ of nodes is given by
\begin{equation}
	\Omega = \begin{cases}
            \Omega_{s}/\Omega_{m}				& \text{if} \ \Omega_{s} \leq \Omega_{m} \\
            1 										&	\text{otherwise}
        	\end{cases}
\end{equation}
where $\Omega_{s} = \sum\limits_{n\in S}{p_n}$. The main idea here is that $0 \leq \Omega_{m} \leq 1$ quantifies the amount of damage the network will tolerate before it collapses, while $\Omega \to 1$ quickly when key nodes for the integrity of the network are removed~\cite{grassia2021machine}~(Fig.~\ref{fig:FigRecovery}). 

Lastly, if robust design and early-warning indicators fail and network collapse is already well underway, then adaptive responses and repair stand as the last two remaining options, as studied in~\cite{Majdandzic2016optimal} for interacting financial networks. Moreover, in~\cite{Sun2020optimal} Sun et al. used optimal control theory and reinforcement learning to determine optimal maintenance protocols to offset aging in complex networks. Most recently, Sanhedrai et al.~\cite{sanhedrai2022reviving} developed a two-step recovery scheme, involving in the first place a topological reconstruction, and secondly dynamic interventions to revive a failed network by means of judiciously applied microscopic interventions~(Fig.~\ref{fig:FigRecovery}).

Taken together, the above reviewed protocols present the most commonly used means of preventing and reacting to network collapse, and they also likely form the most fertile grounds for future developments along these lines.

\section{Conclusions and future research directions}

The analysis of the spectrum of possible responses of a system to external perturbations or internal failures for practical applications comes with several challenges, but also with many opportunities in different disciplines.

At a societal level, our globalized economy relies on an interconnected, multilayer network of economic relationships. Similarly, human movements and communications are no longer geographically localized, and form intricate webs. The robustness of our interdependent society is strictly related to the robustness of its underlying networks: the case of COVID-19 pandemic, as well as the escalation following the political relationships between Russia and Ukraine, have shown that each localized perturbation has the potential to become a systemic problem, such as dysfunction or abnormal function. This fact is a signature of critical behaviour in complex systems. 

As discussed in this review, the problems connected to network dismantling are many, ranging from theoretical challenges to problems concerning the development of algorithms able to dismantle large networks in a reasonable time. Although much work has been done so far, in each of the above points there are still many open questions that require further research efforts.

One aspect that needs to be investigated is related to structure and function, and their interplay. 
Most of the dismantling methods present in literature are based on attacks whose purpose is to strongly decrease the size of the largest connected components. While, in many cases, the size of the largest connected component is somehow linked to the
efficient functioning of a network, in general this is not always true.
For instance, a task-dependent dismantling policy on some networks could produce more devastating effects than one based only on topological criteria.
As an example, the knowledge of the dynamics governing a power-grid network could suggest a more effective attack policy based on the removal of those nodes/arcs most critical for the network stability.
In such a situation, many of the methods present in the literature are inapplicable and there is a strong 
need for new theoretical and computational techniques capable of functional dismantling networks, while accounting for the aspects connected with the dynamics on top of them~\cite{ghavasieh2021unraveling,ghavasieh2023dismantling}.

To this aim, dismantling approaches based on the latent geometry of complex networks~\cite{boguna2021network} also 
 provide a promising research direction. While methods based on machine learning already work in a latent space, there is still no in-depth investigation on how the geometry of a network is linked to its robustness.
 This knowledge could lead to the development of completely new type of algorithms for both structural and functional network dismantling, but also for the improvement of the robustness and resilience of networks.
In fact, while AI--based techniques, such as machine learning, have had a large impact on many fields like natural language, video and image processing, with incredibly popular applications like chat bots and image generation, such impact has been, so far, more limited on networks.
However, it is inevitable that such advancements --- that have roots in the understanding of the techniques, of the algorithms, and of the hardware --- will find their way to many network-related tasks.

Overall, understanding how systems react and adapt to external shocks is an overarching objective for the future.
Research at the edge of statistical physics, network science and machine learning is expected to play a fundamental role for this purpose. Potential applications span from the cellular scale -- where molecular and systems biology need robust techniques to assess the potential impact of genetic or pharmaceutical interventions -- to epidemiology and economics, where shocks have the potential to become systemic and affect the structure and functioning of our society.

\subsection*{Code availability}

The code used for the performance comparisons can be found in the repository \url{https://github.com/NetworkDismantling/review}.


\section{Supplementary Information}

\colorlet{tableheadcolor}{gray!25} 
\newcommand{\headcol}{\rowcolor{tableheadcolor}} %
\colorlet{tablerowcolor}{gray!10} 
\newcommand{\rowcol}{\rowcolor{tablerowcolor}} %

\begin{table}[htpb!]
\begin{adjustbox}{max width=\textwidth}
\begin{tabular}{clll}
\toprule
\multicolumn{1}{l}{} & Network & Name & References \\
\midrule
\cellcolor[HTML]{BF0041} & advogato & Advogato trust network & \cite{konect:massa09} \\
\cellcolor[HTML]{BF0041} & arenas-meta & C. elegans & \cite{konect:duch05} \\
\cellcolor[HTML]{BF0041} & ARK201012\_LCC & CAIDA ARK (Dec 2010) (LCC) & \cite{CAIDA_ARK} \\
\cellcolor[HTML]{BF0041} & cfinder-google & Google.com internal & \cite{konect:palla07} \\
\cellcolor[HTML]{BF0041} & cit-HepPh & arXiv hep-ph & \cite{konect:leskovec107} \\
\cellcolor[HTML]{BF0041} & citeseer & CiteSeer & \cite{b524} \\
\cellcolor[HTML]{BF0041} & com-amazon & Amazon (MDS) & \cite{konect:leskovec2012} \\
\cellcolor[HTML]{BF0041} & com-dblp & DBLP co-authorship & \cite{konect:leskovec2012} \\
\multirow{-9}{*}{\cellcolor[HTML]{BF0041}\textbf{\begin{tabular}[c]{@{}c@{}}B\\  I\\  O\end{tabular}}} & com-youtube & YouTube friendships & \cite{konect:leskovec2012} \\
\cellcolor[HTML]{2A6099} & corruption & Corruption Scandals & \cite{10.1093/comnet/cny002} \\
\cellcolor[HTML]{2A6099} & dblp-cite & DBLP citation & \cite{konect:DBLP} \\
\cellcolor[HTML]{2A6099} & digg-friends & Digg friends & \cite{konect:digg} \\
\cellcolor[HTML]{2A6099} & dimacs10-celegansneural & C. elegans (neural) & \cite{konect:duncan98} \\
\cellcolor[HTML]{2A6099} & dimacs10-polblogs & Political blogs (LCC) & \cite{konect:adamic2005} \\
\cellcolor[HTML]{2A6099} & dimacs9-COL & DIMACS Implementation Challenge (Colorado) & \cite{konect} \\
\cellcolor[HTML]{2A6099} & dimacs9-NY & DIMACS Implementation Challenge (New York City) & \cite{konect} \\
\cellcolor[HTML]{2A6099} & douban & Douban social network & \cite{konect:socialcomputing} \\
\cellcolor[HTML]{2A6099} & econ-wm1 & Economic network WM1 & \cite{nr} \\
\cellcolor[HTML]{2A6099} & ego-twitter & Twitter lists & \cite{konect:McAuley2012} \\
\cellcolor[HTML]{2A6099} & email-EuAll & EU institution email & \cite{konect:leskovec107} \\
\cellcolor[HTML]{2A6099} & eu-powergrid & SciGRID Power Europe & \cite{SciGRIDv0.2} \\
\cellcolor[HTML]{2A6099} & foodweb-baydry & Florida ecosystem dry & \cite{ulanowicz1999network} \\
\cellcolor[HTML]{2A6099} & foodweb-baywet & Florida ecosystem wet & \cite{ulanowicz1999network} \\
\cellcolor[HTML]{2A6099} & gridkit-eupowergrid & GridKit Power Europe & \cite{wiegmans_2016} \\
\cellcolor[HTML]{2A6099} & gridkit-north\_america & GridKit Power North-America & \cite{wiegmans_2016} \\
\cellcolor[HTML]{2A6099} & hyves & Hyves social network & \cite{konect:socialcomputing} \\
\multirow{-18}{*}{\cellcolor[HTML]{2A6099}\textbf{\begin{tabular}[c]{@{}c@{}}I\\  N\\  F\\  O\\  R\\  M\\  A\\  T\\  I\\  O\\  N\end{tabular}}} & inf-USAir97 & US Air lines (1997) & \cite{nr,pajek_repo} \\
\cellcolor[HTML]{468A1A} & internet-topology & Internet (AS) topology & \cite{konect:zhang05} \\
\cellcolor[HTML]{468A1A} & librec-ciaodvd-trust & CiaoDVD trust network & \cite{konect:ciaodvd} \\
\cellcolor[HTML]{468A1A} & librec-filmtrust-trust & FilmTrust trust network & \cite{konect:filmtrust} \\
\cellcolor[HTML]{468A1A} & linux & Linux source code files & \cite{konect} \\
\cellcolor[HTML]{468A1A} & loc-brightkite & Brightkite friendships & \cite{konect:cho2011} \\
\cellcolor[HTML]{468A1A} & loc-gowalla & Gowalla friendships & \cite{konect:cho2011} \\
\cellcolor[HTML]{468A1A} & london\_transport\_aggr & Aggregated London Transportation network & \cite{DeDomenico8351} \\
\cellcolor[HTML]{468A1A} & maayan-figeys & Human protein (Figeys) & \cite{konect:figeys} \\
\cellcolor[HTML]{468A1A} & maayan-foodweb & Little Rock Lake food web & \cite{konect:little-rock-lake} \\
\cellcolor[HTML]{468A1A} & maayan-Stelzl & Human protein (Stelzl) & \cite{konect:stelzl} \\
\cellcolor[HTML]{468A1A} & maayan-vidal & Human protein (Vidal) & \cite{konect:proteome} \\
\cellcolor[HTML]{468A1A} & moreno\_crime\_projected & Crime (projection) & \cite{konect} \\
\cellcolor[HTML]{468A1A} & moreno\_propro & Protein & \cite{konect:coulomb2005,konect:han2005,konect:stumpf2005} \\
\cellcolor[HTML]{468A1A} & moreno\_train & Train bombing terrorist contacts & \cite{konect:hayes} \\
\cellcolor[HTML]{468A1A} & munmun\_digg\_reply\_LCC & Digg social network replies (LCC) & \cite{konect:choudhury09} \\
\cellcolor[HTML]{468A1A} & munmun\_twitter\_social & Twitter follows (ICWSM) & \cite{konect:choudhury10} \\
\cellcolor[HTML]{468A1A} & opsahl-openflights & OpenFlights & \cite{konect:opsahl2010b} \\
\cellcolor[HTML]{468A1A} & opsahl-powergrid & US power grid & \cite{konect:duncan98} \\
\cellcolor[HTML]{468A1A} & opsahl-ucsocial & UC Irvine messages & \cite{konect:opsahl09} \\
\cellcolor[HTML]{468A1A} & oregon2\_010526 & Autonomous systems Oregon-2 & \cite{10.1145/1081870.1081893} \\
\cellcolor[HTML]{468A1A} & p2p-Gnutella06 & Gnutella P2P, August 8 2002 & \cite{snapnets} \\
\cellcolor[HTML]{468A1A} & p2p-Gnutella31 & Gnutella P2P, August 31 2002 & \cite{konect:ripeanu02} \\
\cellcolor[HTML]{468A1A} & pajek-erdos & Erdős co-authorship network & \cite{pajek_repo} \\
\cellcolor[HTML]{468A1A} & petster-cat-household & Catster households online social & \cite{konect} \\
\multirow{-25}{*}{\cellcolor[HTML]{468A1A}\textbf{\begin{tabular}[c]{@{}c@{}}S\\  O\\  C\\  I\\  A\\  L\end{tabular}}} & petster-catdog-household & Catster/Dogster familylinks (LCC) & \cite{konect} \\
\cellcolor[HTML]{FFFF00} & petster-hamster & Hamsterster full & \cite{konect} \\
\cellcolor[HTML]{FFFF00} & power-eris1176 & Power network problem & \cite{nr} \\
\cellcolor[HTML]{FFFF00} & roads-california & California Road Network & \cite{10.1007/11535331_16} \\
\cellcolor[HTML]{FFFF00} & roads-northamerica & North-America Road Network & \cite{dcws} \\
\cellcolor[HTML]{FFFF00} & roads-sanfrancisco & San Francisco Road Network & \cite{brinkhoff2002framework} \\
\cellcolor[HTML]{FFFF00} & route-views & Autonomous systems AS-733 & \cite{konect:leskovec107} \\
\cellcolor[HTML]{FFFF00} & slashdot-threads & Slashdot threads & \cite{konect:slashdot-threads} \\
\cellcolor[HTML]{FFFF00} & slashdot-zoo & Slashdot Zoo & \cite{kunegis:slashdot-zoo} \\
\cellcolor[HTML]{FFFF00} & soc-Epinions1 & Epinions.com trust network & \cite{konect:richardson2003} \\
\cellcolor[HTML]{FFFF00} & subelj\_jdk & JDK dependency network & \cite{konect} \\
\cellcolor[HTML]{FFFF00} & subelj\_jung-j & JUNG and Javax dependency network & \cite{konect:dependency1} \\
\cellcolor[HTML]{FFFF00} & tech-RL-caida & Internet router network & \cite{nr} \\
\cellcolor[HTML]{FFFF00} & twitter\_LCC & Twitter users (LCC) & \cite{morone2015influence} \\
\cellcolor[HTML]{FFFF00} & web-EPA & Pages linking to epa.gov & \cite{nr} \\
\cellcolor[HTML]{FFFF00} & web-NotreDame & Notre Dame web pages & \cite{konect:albert1999} \\
\cellcolor[HTML]{FFFF00} & web-Stanford & Stanford University web pages & \cite{konect:snap} \\
\cellcolor[HTML]{FFFF00} & web-webbase-2001 & Web network & \cite{nr} \\
\multirow{-18}{*}{\cellcolor[HTML]{FFFF00}\textbf{\begin{tabular}[c]{@{}c@{}}T\\  E\\  C\\  H\\  N\\  O\\  L\\  O\\  G\\  I\\  C\\  A\\  L\end{tabular}}} & wordnet-words & WordNet lexical network & \cite{konect:fellbaum98} \\
\bottomrule
\end{tabular}
\end{adjustbox}
\caption{{\scriptsize Real-world test networks: brief description and references.}}
\label{t:test-networks-list}
\end{table}

\begin{table}[htbp!]
\begin{adjustbox}{max width=\textwidth}
\centering

\begin{tabular}{@{}cllllllllll@{}}
\toprule
\multicolumn{1}{l}{\textbf{}} & \multicolumn{1}{c}{} & \multicolumn{1}{c}{} & \multicolumn{1}{c}{} & \multicolumn{2}{c}{Degree} & \multicolumn{1}{c}{} & \multicolumn{2}{c}{Clustering} & \multicolumn{2}{c}{$k$--core num.} \\
\multicolumn{1}{l}{\textbf{}} & Network & \multicolumn{1}{c}{$\lvert V \rvert$} & \multicolumn{1}{c}{$\lvert E \rvert$} & \multicolumn{1}{c}{$\langle d \rangle$} & \multicolumn{1}{c}{Assort.} & \multicolumn{1}{c}{Density} & \multicolumn{1}{c}{T} & \multicolumn{1}{c}{$\langle \mathrm{LCC} \rangle$} & \multicolumn{1}{c}{$k_{max}$} & \multicolumn{1}{c}{$\langle k \rangle$} \\ \midrule
\cellcolor[HTML]{BF0041} & arenas-meta & 453 & 2.0K & 9.0 & -0.224 & 0.01993 & 0.124 & 0.646 & 10 & 4.93 \\
\cellcolor[HTML]{BF0041} & dimacs10-celegansneural & 297 & 2.1K & 14.5 & -0.303 & 0.04887 & 0.181 & 0.292 & 10 & 7.98 \\
\cellcolor[HTML]{BF0041} & foodweb-baydry & 128 & 2.1K & 32.9 & -0.134 & 0.25910 & 0.314 & 0.335 & 24 & 20.96 \\
\cellcolor[HTML]{BF0041} & foodweb-baywet & 128 & 2.1K & 32.4 & -0.159 & 0.25529 & 0.312 & 0.335 & 23 & 20.30 \\
\cellcolor[HTML]{BF0041} & maayan-figeys & 2.2K & 6.4K & 5.7 & -0.140 & 0.00257 & 0.008 & 0.040 & 10 & 3.02 \\
\cellcolor[HTML]{BF0041} & maayan-foodweb & 183 & 2.5K & 26.8 & -0.506 & 0.14724 & 0.332 & 0.323 & 24 & 16.84 \\
\cellcolor[HTML]{BF0041} & maayan-Stelzl & 1.7K & 3.2K & 3.7 & -0.038 & 0.00219 & 0.006 & 0.006 & 7 & 2.04 \\
\cellcolor[HTML]{BF0041} & maayan-vidal & 3.1K & 6.7K & 4.3 & -0.073 & 0.00137 & 0.035 & 0.064 & 7 & 2.51 \\
\multirow{-9}{*}{\cellcolor[HTML]{BF0041}\textbf{\begin{tabular}[c]{@{}c@{}}B\\  I\\  O  \end{tabular}}} & moreno\_propro & 1.9K & 2.3K & 2.4 & -0.108 & 0.00130 & 0.055 & 0.067 & 5 & 1.48 \\
\cellcolor[HTML]{2A6099} & cfinder-google & 15.8K & 149.5K & 19.0 & -0.083 & 0.00120 & 0.013 & 0.518 & 102 & 11.83 \\
\cellcolor[HTML]{2A6099} & cit-HepPh & 34.5K & 420.9K & 24.4 & 0.119 & 0.00071 & 0.146 & 0.285 & 30 & 12.74 \\
\cellcolor[HTML]{2A6099} & citeseer & 384.4K & 1.7M & 9.1 & -0.021 & 0.00002 & 0.050 & 0.179 & 16 & 4.83 \\
\cellcolor[HTML]{2A6099} & com-amazon & 334.9K & 925.9K & 5.5 & -0.079 & 0.00002 & 0.205 & 0.397 & 6 & 3.35 \\
\cellcolor[HTML]{2A6099} & com-dblp & 317.1K & 1.0M & 6.6 & 0.134 & 0.00002 & 0.306 & 0.632 & 113 & 4.22 \\
\cellcolor[HTML]{2A6099} & dblp-cite & 12.6K & 49.6K & 7.9 & -0.042 & 0.00063 & 0.062 & 0.117 & 12 & 4.11 \\
\cellcolor[HTML]{2A6099} & dimacs10-polblogs & 1.2K & 16.7K & 27.3 & -0.256 & 0.02233 & 0.226 & 0.320 & 36 & 14.79 \\
\cellcolor[HTML]{2A6099} & econ-wm1 & 260 & 2.6K & 19.6 & 0.267 & 0.07585 & 0.567 & 0.267 & 33 & 12.61 \\
\cellcolor[HTML]{2A6099} & linux & 30.8K & 213.7K & 13.9 & -0.034 & 0.00045 & 0.003 & 0.128 & 23 & 7.13 \\
\cellcolor[HTML]{2A6099} & p2p-Gnutella06 & 8.7K & 31.5K & 7.2 & 0.019 & 0.00083 & 0.008 & 0.007 & 9 & 3.98 \\
\cellcolor[HTML]{2A6099} & p2p-Gnutella31 & 62.6K & 147.9K & 4.7 & -0.246 & 0.00008 & 0.004 & 0.005 & 6 & 2.52 \\
\cellcolor[HTML]{2A6099} & subelj\_jdk & 6.4K & 53.7K & 16.7 & -0.126 & 0.00259 & 0.011 & 0.671 & 65 & 9.20 \\
\cellcolor[HTML]{2A6099} & subelj\_jung-j & 6.1K & 50.3K & 16.4 & -0.129 & 0.00269 & 0.011 & 0.675 & 65 & 9.11 \\
\cellcolor[HTML]{2A6099} & web-EPA & 4.3K & 8.9K & 4.2 & -0.251 & 0.00098 & 0.012 & 0.071 & 6 & 2.24 \\
\cellcolor[HTML]{2A6099} & web-NotreDame & 325.7K & 1.1M & 6.9 & -0.008 & 0.00002 & 0.088 & 0.235 & 156 & 4.48 \\
\cellcolor[HTML]{2A6099} & web-Stanford & 281.9K & 2.0M & 14.1 & -0.012 & 0.00005 & 0.009 & 0.598 & 71 & 7.91 \\
\cellcolor[HTML]{2A6099} & web-webbase-2001 & 16.1K & 25.6K & 3.2 & -0.050 & 0.00020 & 0.025 & 0.224 & 32 & 1.90 \\
\multirow{-18}{*}{\cellcolor[HTML]{2A6099}\textbf{\begin{tabular}[c]{@{}c@{}}I\\  N\\  F\\  O\\  R\\  M\\  A\\  T\\  I\\  O\\  N\end{tabular}}} & wordnet-words & 146.0K & 657.0K & 9.0 & -0.009 & 0.00006 & 0.096 & 0.602 & 31 & 5.29 \\
\cellcolor[HTML]{468A1A} & advogato & 6.5K & 43.3K & 13.2 & 0.043 & 0.00202 & 0.092 & 0.195 & 26 & 7.50 \\
\cellcolor[HTML]{468A1A} & com-youtube & 1.1M & 3.0M & 5.3 & -0.009 & 0.00000 & 0.006 & 0.081 & 51 & 2.70 \\
\cellcolor[HTML]{468A1A} & corruption & 309 & 3.3K & 21.2 & 0.837 & 0.06895 & 0.847 & 0.929 & 46 & 19.51 \\
\cellcolor[HTML]{468A1A} & digg-friends & 279.6K & 1.5M & 11.1 & 0.001 & 0.00004 & 0.061 & 0.092 & 176 & 5.69 \\
\cellcolor[HTML]{468A1A} & douban & 154.9K & 327.2K & 4.2 & -0.264 & 0.00003 & 0.010 & 0.016 & 15 & 2.15 \\
\cellcolor[HTML]{468A1A} & ego-twitter & 23.4K & 32.8K & 2.8 & -0.192 & 0.00012 & 0.021 & 0.069 & 10 & 1.45 \\
\cellcolor[HTML]{468A1A} & email-EuAll & 265.2K & 365.6K & 2.8 & -0.026 & 0.00001 & 0.004 & 0.067 & 39 & 1.45 \\
\cellcolor[HTML]{468A1A} & hyves & 1.4M & 2.8M & 4.0 & -0.011 & 0.00000 & 0.002 & 0.045 & 39 & 2.04 \\
\cellcolor[HTML]{468A1A} & librec-ciaodvd-trust & 4.7K & 33.1K & 14.2 & 0.186 & 0.00305 & 0.180 & 0.118 & 43 & 7.52 \\
\cellcolor[HTML]{468A1A} & librec-filmtrust-trust & 874 & 1.3K & 3.0 & 0.105 & 0.00343 & 0.192 & 0.162 & 8 & 1.79 \\
\cellcolor[HTML]{468A1A} & loc-brightkite & 58.2K & 214.1K & 7.4 & 0.049 & 0.00013 & 0.111 & 0.172 & 52 & 3.89 \\
\cellcolor[HTML]{468A1A} & loc-gowalla & 196.6K & 950.3K & 9.7 & -0.022 & 0.00005 & 0.023 & 0.237 & 51 & 5.06 \\
\cellcolor[HTML]{468A1A} & moreno\_crime\_projected & 754 & 2.1K & 5.6 & 0.147 & 0.00749 & 0.581 & 0.726 & 17 & 4.62 \\
\cellcolor[HTML]{468A1A} & moreno\_train & 64 & 243 & 7.6 & 0.068 & 0.12054 & 0.561 & 0.622 & 10 & 5.61 \\
\cellcolor[HTML]{468A1A} & munmun\_digg\_reply\_LCC & 29.7K & 84.8K & 5.7 & -0.015 & 0.00019 & 0.006 & 0.005 & 9 & 2.96 \\
\cellcolor[HTML]{468A1A} & munmun\_twitter\_social & 465.0K & 833.5K & 3.6 & -0.250 & 0.00001 & 0.001 & 0.015 & 30 & 1.81 \\
\cellcolor[HTML]{468A1A} & opsahl-ucsocial & 1.9K & 13.8K & 14.6 & -0.186 & 0.00768 & 0.057 & 0.109 & 20 & 7.77 \\
\cellcolor[HTML]{468A1A} & pajek-erdos & 6.9K & 11.9K & 3.4 & -0.077 & 0.00049 & 0.036 & 0.124 & 10 & 1.75 \\
\cellcolor[HTML]{468A1A} & petster-cat-household & 68.9K & 494.9K & 14.4 & -0.036 & 0.00021 & 0.004 & 0.475 & 79 & 7.32 \\
\cellcolor[HTML]{468A1A} & petster-catdog-household & 324.9K & 2.6M & 16.3 & -0.039 & 0.00005 & 0.007 & 0.245 & 99 & 8.25 \\
\cellcolor[HTML]{468A1A} & petster-hamster & 2.4K & 16.6K & 13.7 & 0.256 & 0.00565 & 0.231 & 0.538 & 24 & 8.08 \\
\cellcolor[HTML]{468A1A} & slashdot-threads & 51.1K & 117.4K & 4.6 & -0.027 & 0.00009 & 0.006 & 0.020 & 15 & 2.37 \\
\cellcolor[HTML]{468A1A} & slashdot-zoo & 79.1K & 467.7K & 11.8 & -0.026 & 0.00015 & 0.024 & 0.058 & 54 & 6.04 \\
\cellcolor[HTML]{468A1A} & soc-Epinions1 & 75.9K & 405.7K & 10.7 & 0.023 & 0.00014 & 0.066 & 0.138 & 67 & 5.48 \\
\multirow{-25}{*}{\cellcolor[HTML]{468A1A}\textbf{\begin{tabular}[c]{@{}c@{}}S\\  O\\  C\\  I\\  A\\  L\end{tabular}}} & twitter\_LCC & 532.3K & 694.6K & 2.6 & -0.013 & 0.00000 & 0.001 & 0.014 & 8 & 1.43 \\
\cellcolor[HTML]{FFFF00} & ARK201012\_LCC & 29.3K & 78.1K & 5.3 & -0.037 & 0.00018 & 0.016 & 0.382 & 33 & 2.71 \\
\cellcolor[HTML]{FFFF00} & dimacs9-COL & 435.7K & 521.2K & 2.4 & 0.003 & 0.00001 & 0.025 & 0.017 & 3 & 1.70 \\
\cellcolor[HTML]{FFFF00} & dimacs9-NY & 264.3K & 365.1K & 2.8 & 0.154 & 0.00001 & 0.025 & 0.021 & 3 & 1.81 \\
\cellcolor[HTML]{FFFF00} & eu-powergrid & 1.5K & 1.8K & 2.5 & -0.077 & 0.00169 & 0.119 & 0.126 & 2 & 1.58 \\
\cellcolor[HTML]{FFFF00} & gridkit-eupowergrid & 13.8K & 17.3K & 2.5 & -0.016 & 0.00018 & 0.100 & 0.089 & 3 & 1.68 \\
\cellcolor[HTML]{FFFF00} & gridkit-north\_america & 16.2K & 20.2K & 2.5 & 0.021 & 0.00015 & 0.102 & 0.085 & 3 & 1.71 \\
\cellcolor[HTML]{FFFF00} & inf-USAir97 & 332 & 2.1K & 12.8 & -0.126 & 0.03869 & 0.396 & 0.625 & 26 & 7.67 \\
\cellcolor[HTML]{FFFF00} & internet-topology & 34.8K & 107.7K & 6.2 & -0.043 & 0.00018 & 0.049 & 0.289 & 63 & 3.21 \\
\cellcolor[HTML]{FFFF00} & london\_transport\_aggr & 369 & 430 & 2.3 & 0.143 & 0.00633 & 0.051 & 0.029 & 2 & 1.60 \\
\cellcolor[HTML]{FFFF00} & opsahl-openflights & 2.9K & 15.7K & 10.7 & 0.147 & 0.00363 & 0.255 & 0.453 & 28 & 5.73 \\
\cellcolor[HTML]{FFFF00} & opsahl-powergrid & 4.9K & 6.6K & 2.7 & -0.073 & 0.00054 & 0.103 & 0.080 & 5 & 1.74 \\
\cellcolor[HTML]{FFFF00} & oregon2\_010526 & 11.5K & 32.7K & 5.7 & -0.068 & 0.00050 & 0.037 & 0.352 & 31 & 2.98 \\
\cellcolor[HTML]{FFFF00} & power-eris1176 & 1.2K & 9.9K & 16.8 & 0.957 & 0.01428 & 0.940 & 0.432 & 81 & 15.68 \\
\cellcolor[HTML]{FFFF00} & roads-california & 21.0K & 21.7K & 2.1 & -0.003 & 0.00010 & 0.000 & 0.000 & 2 & 1.89 \\
\cellcolor[HTML]{FFFF00} & roads-northamerica & 175.8K & 179.1K & 2.0 & 0.004 & 0.00001 & 0.001 & 0.000 & 2 & 1.73 \\
\cellcolor[HTML]{FFFF00} & roads-sanfrancisco & 175.0K & 221.8K & 2.5 & -0.001 & 0.00001 & 0.028 & 0.020 & 3 & 1.72 \\
\cellcolor[HTML]{FFFF00} & route-views & 6.5K & 13.9K & 4.3 & -0.051 & 0.00066 & 0.010 & 0.252 & 12 & 2.41 \\
\multirow{-18}{*}{\cellcolor[HTML]{FFFF00}\textbf{\begin{tabular}[c]{@{}c@{}}T\\  E\\  C\\  H\\  N\\  O\\  L\\  O\\  G\\  I\\  C\\  A\\  L\end{tabular}}} & tech-RL-caida & 190.9K & 607.6K & 6.4 & 0.137 & 0.00003 & 0.061 & 0.158 & 32 & 3.45 \\ \cmidrule(l){2-11} 
\end{tabular}

\end{adjustbox}
\caption{{\scriptsize Real-world test networks: topological measures. Columns $\langle d \rangle$ and ``assort.'' stand for degree average and assortativity respectively, ``$T$'' for transitivity, and ``$\langle \mathrm{LCC} \rangle$'' for average local clustering coefficient.}}
\label{t:network_stats}
\end{table}

\begin{table}[htbp!]
\begin{adjustbox}{max width=\textwidth}
\begin{tabular}{@{}ll|lllllll|llllllll@{}}
\toprule
Network & $\lvert V \rvert$ & AD & BC & EI $\sigma_1$ & FINDER & GDM & GND & MS & PR & CI $\ell-2$ & CoreHD & GDM +R & GND +R & MS +R \\
\midrule
\multirow{3}{*}{CM} & 100K & 110.5 & 142.3 & 111.4 & 108.2 & 114.5 & 131.1 & 106.0 & 125.0 & 100.0 & 106.7 & 108.1 & 121.4 & 105.6 \\
    & 10K & 110.4 & 135.1 & 109.6 & 108.1 & 112.8 & 142.9 & 106.4 & 124.6 & 100.0 & 107.1 & 107.9 & 124.1 & 106.0 \\
    & 1K & 110.5 & 127.9 & 109.2 & 108.4 & 111.5 & 133.3 & 107.2 & 123.6 & 100.0 & 107.8 & 108.0 & 121.3 & 106.9 \\
\multirow{3}{*}{ER} & 100K & 111.3 & 135.2 & 109.0 & 109.6 & 114.1 & 125.2 & 107.2 & 126.0 & 100.0 & 108.9 & 110.3 & 122.3 & 107.0 \\
    & 10K & 111.3 & 133.4 & 108.8 & 109.5 & 114.1 & 125.0 & 107.2 & 126.0 & 100.0 & 109.1 & 110.3 & 122.1 & 107.0 \\
    & 1K & 110.9 & 128.9 & 108.5 & 109.0 & 113.5 & 121.9 & 107.3 & 124.9 & 100.0 & 108.8 & 109.6 & 119.8 & 107.0 \\
\multirow{3}{*}{SBM} & 100K & 108.7 & 120.5 & 106.9 & 108.0 & 112.5 & 109.0 & 106.4 & 118.0 & 100.0 & 107.1 & 108.9 & 108.1 & 105.9 \\
    & 10K & 108.7 & 120.1 & 106.7 & 107.9 & 112.5 & 105.8 & 106.3 & 117.9 & 100.0 & 107.2 & 108.8 & 106.2 & 105.9 \\
    & 1K & 108.4 & 118.9 & 106.5 & 107.4 & 111.8 & 102.7 & 106.1 & 118.0 & 100.0 & 106.7 & 108.2 & 104.6 & 105.5 \\
\midrule
    & Average & 110.1 & 129.1 & 108.5 & 108.4 & 113.0 & 121.9 & 106.7 & 122.7 & 100.0 & 107.7 & 108.9 & 116.6 & 106.3 \\
\bottomrule
\end{tabular}
\end{adjustbox}
\caption{Synthetic network results table. We compare the algorithms on Erd\H{o}s-R\'{e}nyi (ER) networks (average degree $k_{\text{avg}} = 4$), on Configuration Model networks (CM) with power law distribution ($\gamma = 2.5$ and $k_{\text{avg}} = 4$) and on Stochastic Block Model (SBM) networks (group size fixed to $100$, $p_{\text{intra}} = 0.1$ and $p_{\text{inter}} = \frac{5}{\lvert V \rvert}$). Each value is the average on $10$ different model instances. The dismantling target for each method is $10\%$ of the network size. We compute the AUC value by integrating the $\lvert \mathrm{LCC}(x) \rvert /\lvert V \rvert$ --- i.e., the Largest Connected Component's size as a function of the removed nodes --- values using Simpson’s rule. For sake of readability, the AUC values of each network are expressed as the percentage of the value obtained by the best performing method on that network. That is, the lower the AUC, the better. Column ``AD'' stands for Adaptive Degree, ``BC'' for Betweenness Centrality, ``CI'' for Collective Influence, ``EI'' for Explosive Immunization, ``GND'' for Generalized Network Dismantling, ``MS'' for MinSum, and ``PR'' for PageRank. ``+R'' means that the reinsertion phase is performed. We note that CoreHD and CI are compared to other +R algorithms as they include the reinsertion phase.}
\label{t:synth_test_networks_table}
\end{table}

\begin{table}[htbp!]
\begin{adjustbox}{max width=\textwidth}
\begin{tabular}{@{}ll|lllllll|llllll@{}}
\toprule
$\mu$ & $t_1$ & AD & BC & EI $\sigma_1$ & FINDER & GDM & GND & MS & PR & CI $\ell-2$ & CoreHD & GDM +R & GND +R & MS +R \\
\midrule
\multirow{3}{*}{0.1} & 2.2 & 112.9 & 143.8 & 100.0 & 111.4 & 115.7 & 129.9 & 404.1 & 119.8 & 101.8 & 120.3 & 105.6 & 113.5 & 118.8 \\
 & 2.6 & 112.6 & 159.0 & 101.4 & 110.2 & 116.3 & 100.5 & 238.4 & 126.2 & 100.0 & 108.5 & 106.4 & 111.2 & 108.7 \\
 & 3.5 & 144.4 & 214.6 & 131.7 & 140.7 & 149.0 & 100.0 & 216.3 & 167.9 & 125.9 & 136.4 & 137.1 & 123.6 & 136.2 \\
\multirow{3}{*}{0.2} & 2.2 & 109.7 & 132.0 & 100.5 & 108.2 & 113.0 & 156.7 & 381.6 & 117.7 & 100.0 & 117.8 & 105.1 & 133.4 & 118.4 \\
 & 2.6 & 112.1 & 151.4 & 103.2 & 109.5 & 115.5 & 125.2 & 232.1 & 125.8 & 100.0 & 108.7 & 107.3 & 130.3 & 110.0 \\
 & 3.5 & 114.0 & 166.0 & 107.4 & 110.9 & 117.7 & 109.0 & 169.5 & 131.3 & 100.0 & 108.8 & 110.1 & 123.1 & 109.8 \\
\multirow{3}{*}{0.3} & 2.2 & 109.4 & 129.4 & 100.3 & 108.1 & 112.4 & 173.3 & 372.8 & 117.6 & 100.0 & 116.5 & 104.9 & 139.3 & 118.1 \\
 & 2.6 & 111.9 & 146.5 & 103.2 & 109.2 & 115.1 & 132.3 & 227.8 & 126.0 & 100.0 & 107.9 & 106.9 & 128.3 & 109.2 \\
 & 3.5 & 114.0 & 164.6 & 107.7 & 110.8 & 117.6 & 116.0 & 169.2 & 131.5 & 100.0 & 108.8 & 110.2 & 124.1 & 110.0 \\
\multirow{3}{*}{0.4} & 2.2 & 109.9 & 129.6 & 100.8 & 108.2 & 112.4 & 179.0 & 365.7 & 117.9 & 100.0 & 115.3 & 105.0 & 135.5 & 117.4 \\
 & 2.6 & 111.9 & 146.3 & 103.3 & 109.0 & 115.0 & 138.1 & 226.4 & 125.9 & 100.0 & 107.8 & 106.9 & 134.6 & 109.7 \\
 & 3.5 & 113.9 & 164.4 & 107.7 & 110.7 & 117.4 & 118.6 & 168.4 & 131.3 & 100.0 & 108.7 & 110.2 & 124.8 & 110.1 \\
\multirow{3}{*}{0.5} & 2.2 & 110.6 & 130.5 & 101.1 & 108.4 & 112.6 & 185.6 & 359.9 & 118.9 & 100.0 & 114.8 & 105.1 & 136.9 & 116.9 \\
 & 2.6 & 112.1 & 146.9 & 103.4 & 109.4 & 115.0 & 144.2 & 224.5 & 126.1 & 100.0 & 107.9 & 107.1 & 135.4 & 109.3 \\
 & 3.5 & 113.8 & 164.3 & 107.6 & 110.5 & 117.3 & 118.7 & 168.0 & 131.6 & 100.0 & 108.6 & 110.1 & 125.6 & 110.0 \\
 \midrule
\multicolumn{2}{l|}{Average} & 114.2 & 152.6 & 105.3 & 111.7 & 117.5 & 135.1 & 261.6 & 127.7 & 101.9 & 113.1 & 109.2 & 128.0 & 114.2 \\
\bottomrule
\end{tabular}
\end{adjustbox}
\caption{Dismantling performance on the Lancichinetti-Fortunato-Radicchi (LFR)~\cite{PhysRevE.78.046110} model. The networks are generated with $N = 2^14$ nodes, degree exponent $\gamma \in \{2.2, 2.6, 3.5\}$, average degree $\langle k \rangle = 6$, and maximum degree $k_{max} = \sqrt{N} = 128$.
We use values of the mixing parameter $\mu \in \{0.1, \dots , 0.5\}$, while the communities are randomly distributed with a size distribution $P(s) \sim s^{-1}$, where $\sqrt{N}$ ($128$) and $5 \times \sqrt{N}$ ($640$) chosen as minimum and maximum sizes, respectively. The parameter $\gamma$ controls the heterogeneity of the degree distribution, i.e., $P(k) \sim k^{-\gamma}$. The mixing parameter $\mu$ controls the strength of correlation between network topology and imposed embedding, as low-$\mu$ values favor connections between pairs of nodes belonging to the same pre-imposed communities. Each result is the average on $10$ instances. The dismantling target for each method is $10\%$ of the network size. We compute the AUC value by integrating the $\lvert \mathrm{LCC}(x) \rvert /\lvert V \rvert$ --- i.e., the Largest Connected Component's size as a function of the removed nodes --- values using Simpson’s rule. For sake of readability, the AUC values of each network are expressed as the percentage of the value obtained by the best performing method on that network. That is, the lower the AUC, the better. Column ``AD'' stands for Adaptive Degree, ``BC'' for Betweenness Centrality, ``CI'' for Collective Influence, ``EI'' for Explosive Immunization, ``GND'' for Generalized Network Dismantling, ``MS'' for MinSum, and ``PR'' for PageRank. ``+R'' means that the reinsertion phase is performed. We note that CoreHD and CI are compared to other +R algorithms as they include the reinsertion phase.}
\label{t:lfr_results}
\end{table}

\bibliography{biblio}


\end{document}